%% file: main.tex
\documentclass[conference,compsoc]{IEEEtran}

\usepackage[T1]{fontenc}
\usepackage[utf8]{inputenc}
\usepackage[nocompress]{cite}
\usepackage{amsmath}
\usepackage{amssymb}
\usepackage{microtype}
\usepackage{xcolor}
\usepackage{graphicx}
\usepackage{enumitem}
\usepackage{needspace}
\setlist{labelindent=0pt,leftmargin=*,align=left,
  topsep=\IEEEiedtopsep,itemsep=0pt,parsep=0pt,partopsep=0pt}
\setlist[enumerate,1]{label=\arabic*)}
\usepackage{xspace}
\usepackage{acro}
\usepackage{tabularx}
\usepackage{booktabs}
\usepackage{multirow}
\usepackage{tikz}
\usepackage[skins,breakable]{tcolorbox}
\usepackage{xurl}
\usepackage[hidelinks]{hyperref}
\usepackage{flushend}
\usetikzlibrary{shapes.geometric, arrows.meta, positioning, calc}

\newcommand{\pp}[1]{\par\noindent\textbf{#1}}
\newcounter{insight}
\renewcommand{\theinsight}{I\arabic{insight}}
\newcommand{\insight}[1]{\par\noindent\refstepcounter{insight}\textbf{\theinsight: #1}}
\newcolumntype{Y}{>{\raggedright\arraybackslash}X}

\newcommand{\sv}[1]{\textsuperscript{\textsc{#1}}}

\definecolor{ClaimBack}{RGB}{255,245,247}
\definecolor{ClaimFrame}{RGB}{214,94,116}
\tcbset{claimstyle/.style={enhanced, breakable, colback=ClaimBack,
    colframe=ClaimFrame, coltitle=black, fonttitle=\bfseries,
    boxrule=0.45pt, arc=2pt, left=6pt, right=6pt, top=5pt, bottom=5pt,
    before skip=6pt, after skip=8pt}}
\newcommand{\auditwarning}{%
  \tikz[x=.78em,y=.70em,baseline=0pt]{%
    \path[draw=black,fill=black!12,line width=.45pt,
      line join=round] (0,0) -- (1,0) -- (.5,.9) -- cycle;
    \node[inner sep=0pt,text=black,font=\sffamily\bfseries\fontsize{5.5}{5.5}\selectfont]
      at (.5,.31) {!};}}
\newtcolorbox{scopeclaimbox}{claimstyle,
  before upper={\noindent\auditwarning\enspace}}
\newcommand{\toolname}{\textsc{A0}\xspace}
\newcommand{\ModelSol}{\texttt{gpt-6.1-sol}\xspace}
\newcommand{\DatasetKF}{\ensuremath{\mathrm{Dataset}_{\mathrm{KF}}^{42}}}
\newcommand{\DatasetITW}{\ensuremath{\mathrm{Dataset}_{\mathrm{ITW}}^{121}}}
\DeclareAcronym{llm}{short = LLM, long = large language model}

\begin{document}

\title{Can AI Agents Find Flaws in Cryptographic Proofs?}

\author{\IEEEauthorblockN{Tiancheng Mai, Qiang Tang, Sri AravindaKrishnan
    Thyagarajan, Ziyue Wang, and Liyi Zhou}
  \IEEEauthorblockA{The University of Sydney}}

\maketitle
\pagestyle{plain}
\thispagestyle{plain}

\begin{abstract}
AI agents perform well in software security, where a candidate finding can be
  validated by machine feedback such as a crash.  Proof auditing offers no such
  oracle: judging a flaw takes expert reasoning about assumptions, games,
  reductions, and probability bounds.  The challenge is turning that judgment
  into an explicitly engineered, human-aligned learning signal.

  We introduce \toolname, a cryptographic-proof auditing system whose reward
  specification is three versioned ``books'': (i) an audit playbook for search;
  (ii) a per-case answerbook of reference flaws; and (iii) a grading playbook
  for assigning credit.  Human-over-the-loop adjudication evolves them as
  experts resolve cases the specification misses.  To our knowledge it is the
  first proof-auditing system with an explicit, human-aligned reward
  specification revised from expert feedback.
\end{abstract}
\acresetall

\section{Introduction}
\label{sec:introduction}

Cryptography is the trust layer of digital infrastructure: encrypted messages,
signatures, and authentication rely on mathematical proofs. A flawed proof
can leave a deployed system without its claimed guarantee for years.

The number of papers posted annually to the Cryptology ePrint Archive rose
from 52 in 2000 to 662 in 2010, 1{,}581 in 2020, and 2{,}797 in 2025:
$54\times$ since 2000 and $1.8\times$ since 2020. Checking security proofs line by line demands
expertise and time that volunteer peer reviewers may lack. Reviewers may
check that a proof ``looks roughly right'' while skipping technical details.
Unsurprisingly, \emph{research output has outpaced the community's verification capacity.}

Fundamentally incorrect proofs can pass review, accumulate citations, and
reach deployment. OCB2, introduced in 2004~\cite{rogaway2004ocb} and incorporated
into ISO/IEC 19772:2009~\cite{iso19772-2009}, was shown in 2018 and 2019 to
admit practical authenticity and confidentiality attacks arising from its
security argument~\cite{inoue2018ocb2,inoue2019ocb2}; the 2009 standard was
withdrawn and replaced~\cite{iso19772-2020}.

Agents powered by \acp{llm} are starting to fill this gap. A frontier LLM found a fatal
cryptographic flaw whose paper was withdrawn~\cite{woodruff2026gemini}, and
an LLM paper-review assistant has been piloted at theory and machine-learning
conferences~\cite{jayaram2026pat}. However, assessing these agents
as proof auditors requires systematic evidence across papers and repeated runs.
We test whether expert-guided audits find more flaws, support them with stronger
evidence, and detect them more consistently.

\pp{Our approach: human-aligned agentic proof auditing.}
We develop \toolname, an agentic system for auditing cryptographic
proofs, with human judgment shaping its audit instructions, reference
findings, and credit criteria.
Experts help build and revise the Answerbook. Its
per-case JSON files identify reference flaws and the evidence required for credit.
The audit playbook guides the auditor through claims and their supporting arguments.
An LLM proposes revisions from grading feedback. We compare each revision with the
current playbook on development cases by replaying audits
(\S\,\ref{sec:playbook-development}). During evaluation, we pool findings
across models and playbook conditions.
We spent approximately 220 person-hours preparing and aligning
references. Source checks and targeted technical review informed which findings
mattered and what evidence earned credit (\S\,\ref{sec:alignment}).

\pp{Research questions.}
We develop the audit playbook using historical cases with
documented proof flaws (RQ1). We then compare audits with ($+P$) and without
($-P$) the playbook on a separate known-flaw benchmark (RQ2) and on recent
papers selected without requiring a known flaw (RQ3).
\begin{itemize}
  \item \textit{RQ1 (Playbook development): How can LLM-assisted
        revision and expert feedback be combined to develop a reusable
        playbook for cryptographic proof auditing?}
  \item \textit{RQ2 (Effectiveness): How does the playbook affect models' ability
        to detect proof flaws on a known-flaw benchmark?}
        We ask how it changes the number of flaws found, the evidence
        provided, and the consistency of findings across models and
        repeated runs.
  \item \textit{RQ3 (In-the-wild audit performance): How does the playbook
        affect audit scores and coverage on papers selected without requiring
        a known flaw?}
\end{itemize}

\pp{Key findings across two datasets.}
\begin{itemize}
  \item \textbf{Known-flaw benchmark}
        (\DatasetKF{}; \S\,\ref{sec:rq2}).
        The benchmark contains 42 cryptographic papers with documented
        proof flaws. Published attacks identify known flaws but may miss others.
        We test five models
        (GPT-5.6/6/6.1 Sol, GLM-5.3-Flash, and MiniMax-M3.1-Flash)
        from different generations and families, with and without the playbook.
        Auditors see only the original papers.
  \item \looseness=-1 \textbf{In-the-wild evaluation}
        (\DatasetITW{}; \S\,\ref{sec:rq3}).
        We chose the model with the highest overall audit score
        on \DatasetKF{} (GPT-6.1 Sol) and evaluated it with and
        without the audit playbook on 121 papers
        accepted at CRYPTO, EUROCRYPT, ASIACRYPT, PKC, and TCC in 2026.
\end{itemize}
The main results below focus on findings classified as Critical
under our severity taxonomy (\S\,\ref{sec:models}). Examples of Low-severity
findings include typos, misnamed assumptions, and incorrect variable usage.
Section~\ref{sec:study} defines the metrics. Here, ``overall'' refers to Critical routes only.

\begin{itemize}
  \item \hypertarget{finding:1}{}\textit{Finding 1 (Overall performance): Our playbook improves overall audit
        scores across models.} All five models score higher on \DatasetKF{}.
        Their equally weighted mean audit score (mean credit per Critical
        route, 0--1) rises by 86.9\% (0.209 to 0.390) at
        $8.1$--$11.7\times$ the estimated inference cost.
        The top scorer, GPT-6.1 Sol, gains 31.5\%
        (0.384 to 0.506; \S\,\ref{sec:rq2-performance}) and 50.7\% on
        \DatasetITW{} (0.433 to 0.652; \S\,\ref{sec:rq3}).
  \item \hypertarget{finding:2}{}\textit{Finding 2 (Coverage): Mean coverage (p0) increases by
        14.0 percentage points for GPT-6.1 Sol on \DatasetKF{},
        while some baseline-credited routes are missed.}
        Across the 81 Critical reference routes in \DatasetKF{},
        GPT-6.1 Sol covers an average of 32.3 without the playbook and 43.7
        with it (39.9\% vs.\ 53.9\%).
        We observe the same pattern across the other four models:
        the playbook credits more routes overall, but some receive credit
        only without it (\S\,\ref{sec:rq2-coverage}). In \DatasetITW{},
        the mean within-round $+P$-only and $-P$-only sets occupy 31.3\% and
        6.8\% of the reference, giving a net gain of 24.5 percentage points
        (\S\,\ref{sec:rq3}).
        On the 98 Critical routes pooled from both conditions,
        GPT-6.1 Sol's mean coverage rises from 45.7 to 69.7 routes per run
        with the playbook (46.6\% to 71.1\%).
        \par
Coverage includes partial credit. In these $+P$/$-P$ comparisons, gains remain positive for every model--dataset pair in every round at the route-specific median threshold (p50; \S\,\ref{sec:study}) and when requiring full rubric credit (Appendix~\ref{app:extended}). Neither endpoint is independent human confirmation.
  \item \hypertarget{finding:3}{}\textit{Finding 3 (Depth): Our playbook deepens justification
        for weaker models (e.g., GPT-5.6 Sol: $+10.7\%$), while GPT-6.1 Sol
        already scores near the rubric ceiling on flaws it finds with and without the playbook.}
        On \DatasetKF{}, we compare justification
        for Critical routes credited both with and without the playbook in the
        same round. GPT-5.6 Sol's mean score rises from 0.821 to 0.908
        ($+10.7\%$). GPT-6.1 Sol already scores close to full credit on
        these routes, with little change from 0.972 to 0.970. Routes credited
        only with the playbook in the same round receive a mean score of
        0.866. For GPT-6.1 Sol, the main gain is therefore broader
        route coverage with substantial supporting
        evidence (\S\,\ref{sec:rq2-depth}).
  \Needspace{3\baselineskip}
  \item \hypertarget{finding:4}{}\textit{Finding 4 (Repeat detection): Our playbook improves
        repeat detection of baseline-credited routes across all evaluated models.}
        We select routes using one without-playbook run and measure
        their recurrence in both remaining runs (\S\,\ref{sec:study}). At p0,
        GPT-5.6 Sol improves from 47.0\% to 88.0\% and GPT-6.1 Sol from
        80.7\% to 87.8\% on \DatasetKF{}; GPT-6.1 improves from
        68.1\% to 78.2\% on \DatasetITW{}. At p0 and p50, every model
        improves in all three overlapping splits
        (Table~\ref{tab:baseline-repeatability}; \S\S\,\ref{sec:rq2-repeat} and~\ref{sec:rq3-runtime}).
  \item \hypertarget{finding:5}{}\textit{Finding 5 (Cost): Mixing models can lower cost
        while preserving or improving overall audit performance.}
        On \DatasetKF{}, three runs of GPT-6.1 Sol
        with the playbook cost about \$13.25 per paper in total and achieve a combined overall
        audit score of 0.639. The combined score averages each Critical route's
        highest single-report credit across the three reports
        (\S\,\ref{sec:study}).
        Across the 42 papers, three runs covered 55 Critical
        routes with the playbook and 39 without it, at about \$31 extra per
        net additional route. To test whether mixing models can improve
        the trade-off between cost and audit performance, we compare different
        model combinations across three audits, all with the playbook.
        GPT-6.1 Sol $+$ MiniMax-M3.1-Flash $\times 2$ costs \$7.94 per paper
        (\$5.32 less), with a score of 0.625 ($-2.3\%$).
        GPT-6.1 Sol $\times 2$ $+$ MiniMax-M3.1-Flash costs \$10.60 per paper
        (\$2.66 less) and scores 0.663 ($+3.8\%$).
        GPT-6.1 Sol $+$ GPT-6 Sol $+$ GLM-5.3-Flash costs \$10.12 per paper
        (\$3.13 less) and scores
        0.665 ($+4.0\%$; \S\,\ref{sec:rq2-cost}). Some mixtures lower cost at a small
        score loss; others improve both metrics.
        We estimate costs at API rates and score the separate reports
        retrospectively (\S\,\ref{sec:study}).
\end{itemize}

\pp{Contributions}\footnote{\url{https://github.com/DisproofAILab/A0-Artifacts}}
\begin{enumerate}
  \item \textbf{Human-guided development of transferable auditing
        playbooks.} To our knowledge, we are the first to systematically
        develop and evaluate human-guided playbooks for auditing the proofs
        in cryptographic research papers. Through expert feedback and
        LLM-assisted revision, we turn judgments about which flaws matter,
        what evidence is sufficient, and how to investigate suspected flaws
        into reusable audit procedures. Comparisons with and without the
        resulting playbook show improvements in overall audit scores across
        five models on \DatasetKF{} and for GPT-6.1 Sol in \DatasetITW{}.
  \item \looseness=-1 \textbf{Datasets.}
        We will release both paper datasets and the Answerbook for the
        known-flaw benchmark. For the in-the-wild dataset, we will keep
        the full Answerbook private and publish a redacted index of its
        reference routes (Section~\ref{sec:responsible-disclosure}).
        Preparing the reference data and conducting technical review took
        an estimated 220 person-hours.
  \item \textbf{The auditing playbook.}
        We will release the audit playbook to help others improve
        proof auditing with their own LLMs and tools and evaluate proof
        auditing on the known-flaw benchmark.
  \item \textbf{In-the-wild auditing.}
        For 121 recent papers, we report audit results and review status. We have
        begun notifying authors about selected concerns
        (Appendix~\ref{app:dataset-reference-details}). Our public index\footnote{\url{https://disproofailab.github.io/A0-Artifacts/}}
        will present all reference routes in the adopted \DatasetITW{} Answerbook
        as potential issues, grouped by potential severity
        (\S\,\ref{sec:responsible-disclosure}).
\end{enumerate}

\begin{scopeclaimbox}
  We chose which proof errors to include, how to classify
  them, which to prioritize, and what evidence earns credit. These choices
  draw on our research and reviewing experience, including service by
  three of us on program committees of CORE A*-ranked security and
  cryptography conferences. They shape both the playbook and its
  evaluation. Other researchers can use the same process with their own
  criteria (\S\S\,\ref{sec:discussion} and~\ref{sec:limitations}).
\end{scopeclaimbox}

\section{Background}
\label{sec:background}

\pp{Provable security claims.}
A paper claiming provable security presents a construction,
a security theorem, and a proof. The theorem typically bounds an efficient
adversary's advantage (otherwise referred to as its chance of a successful attack) in a specified security game under stated assumptions.

\pp{Proof vulnerabilities.}
We use \emph{proof vulnerability} for a substantive gap in this
argument that undermines the justification for the claimed guarantee.
Examples include invalid game hops in hybrid argument-based security proofs, unjustified simulations, and omitted
losses in security bounds; cosmetic issues are excluded. Assessing a gap
requires tracing dependencies across the construction, adversary interfaces,
proof steps, and cited assumptions.

\pp{Peer review.}
Cryptographic papers are vetted by program committees under fixed
deadlines~\cite{dulek2021guide}; Eurocrypt 2025 alone received 624
submissions and accepted 123~\cite{eurocrypt2025press}, and IACR's own
publication proposal warns that review load leaves too little time for long,
proof-heavy papers~\cite{leurent2022iacrpublication}.  Review outcomes are
also noisy~\cite{soneji2022flawed,cortes2021neurips,beygelzimer2023neurips}.
Peer review judges importance and context; it is not a line-by-line proof
check.

\pp{Auditing paper-written proofs.} \looseness=-1
Most cryptographic proofs are published as PDFs of prose, notation,
pseudocode, and cross-references. Auditing
them has no executable oracle: a later attack paper confirms one flaw but
neither enumerates all flaws nor certifies the rest of the proof.
Technical review assesses the evidence for a reported flaw;
an acknowledgment from the paper's authors does not itself establish the
claim's correctness.

\pp{Model, harness, and playbook.} \looseness=-1
The \emph{model} is the \ac{llm} that reads and reasons.
The \emph{harness} is the runtime supplying files, tools, subagents,
context management, and output schemas. A \emph{playbook} specifies
task-specific procedures in text: which claims to inspect, how to
check their dependencies, and what evidence to report.
Procedural failures include inspecting the wrong
theorem, abandoning a promising lead, or omitting supporting reasoning from
the report.

\pp{Evaluating proof-audit findings.} \looseness=-1
Assessing a reported flaw requires checking its reasoning
against the paper's claims and assumptions. LLM-as-a-judge supports grading
at scale with reference answers and task-specific rubrics,
but remains susceptible to bias and reasoning
errors~\cite{zheng2023judging,kim2023prometheus}. Experts can scrutinize
natural-language proofs, but this takes specialist time and requires resolving
disagreements~\cite{ma2025proofgrader,gonzalez2026qedbench}.
QEDBench combines LLMs and expert judgment through LLM-drafted,
expert-refined rubrics for separate expert and LLM
assessments~\cite{gonzalez2026qedbench}. These protocols differ in
who constructs reference criteria, who assigns scores, and which
judgments receive manual technical review.

\section{Problem Setting and Implementation}
\label{sec:models}

\subsection{Audit Task, Taxonomy, and Severity}
\label{sec:audit-task}
\label{sec:taxonomy-severity}

\pp{Task and output.}
The auditor checks a paper's claims and arguments for errors in
Table~\ref{tab:issue-taxonomy}. The report records the outcome, then each finding's
affected claim or component, label, reasoning, and source evidence.

\pp{Evidence and assumptions.}
The audit instructions restrict evidence to the supplied paper
and prohibit consulting external attack papers or Answerbook entries.
We assume honest authors and accidental errors;
documents crafted to manipulate the auditor are beyond the scope of this study.

\pp{Taxonomy.}
Table~\ref{tab:issue-taxonomy} shows how we group the proof
flaws and minor errors within our audit scope. Each finding receives
a \texttt{<bucket><class>} identifier. The five main buckets cover assumptions,
adversarial models, definitions, constructions, and proofs. Residual classes cover
other issues within each bucket, or globally across or outside them.
\begin{table}[tb]
  \centering
  \caption{Taxonomy of \toolname proof-audit findings.
    Superscripts give severity (L = Low, M = Medium, C = Critical, U = Unknown). For
    each bucket 1--5, suffixes x, y and z denote bucket-specific residuals at L, M and
    C, respectively.}
  \label{tab:issue-taxonomy}
  \footnotesize
  \begin{tabularx}{\columnwidth}{@{}>{\raggedright\arraybackslash}p{0.17\columnwidth}@{\hspace{0.015\columnwidth}}Y@{}}
    \hline
    \textbf{Bucket}   & \textbf{Taxonomy IDs}                                  \\
    \hline
    Assumption        &
    \texttt{1a}\sv{l}: misnamed assumption\par
    \texttt{1b}\sv{m}: incorrect or insufficient parameter choices\par
    \texttt{1c}\sv{l}: incorrect write-up of a correctly named assumption\par
    \texttt{1d}\sv{c}: missing assumption \\
    \hline
    Adversarial model &
    \texttt{2a}\sv{m}: incorrect adversary computational power\par
    \texttt{2b}\sv{c}: incorrect or unsupported communication model\par
    \texttt{2c}\sv{c}: incorrect or unsupported corruption timing\par
    \texttt{2d}\sv{m}: missing corruptible entities \\
    \hline
    Definition        &
    \texttt{3a}\sv{l}: misnamed definition; \texttt{3b}\sv{m}: missing oracle access\par
    \texttt{3c}\sv{m}: incorrect winning condition\par
    \texttt{3d}\sv{c}: missing information, computation, or verification\par
    \texttt{3e}\sv{c}: excessive leakage to the adversary\par
    \texttt{3f}\sv{m}: incorrect computational-power requirement\par
    \texttt{3g}\sv{l}: incorrect quantifiers \\
    \hline
    Construction      &
    \texttt{4a}\sv{c}: missing computation\par
    \texttt{4b}\sv{m}: redundant or unnecessary computation\par
    \texttt{4c}\sv{l}: incorrect variable usage\par
    \texttt{4d}\sv{l}: invalid or malformed computation \\
    \hline
    Proof             &
    \texttt{5a}\sv{l}: missing words or sentences; \texttt{5b}\sv{l}: alphanumeric typos\par
    \texttt{5c}\sv{m}: incorrect theorem, lemma, fact, proposition, or claim\par
    \texttt{5d}\sv{c}: missing lemma(s), fact(s), proposition(s), or claims\par
    \texttt{5e}\sv{m}: redundant lemma(s), fact(s), proposition(s), or claims\par
    \texttt{5f}\sv{c}: faulty computational-resource accounting \\
    \hline
    Global residual   &
    \texttt{6a}\sv{l}/\texttt{6b}\sv{m}/\texttt{6c}\sv{c}/\texttt{6u}\sv{u}: unmodeled
    provable-security issues that do not fit a standard bucket or are
    inherently cross-bucket                                                    \\
    \hline
\end{tabularx}
\end{table}

\pp{Severity.}
No single severity scale captures every aspect of an error in a cryptographic proof, and researchers may reasonably differ in which consequences they prioritize.
Table~\ref{tab:issue-taxonomy} records the severity we assign to each flaw class for this study.
Grading uses the severity recorded for each reference route in the adopted Answerbook (Section~\ref{sec:grading}), and our playbook development and principal evaluation prioritize Critical routes.
\S\,\ref{sec:discussion} discusses how other groups can adapt these choices.
The mathematical validity of a reported error is assessed separately from its severity.

\Needspace{5\baselineskip}
\subsection{Auditor Implementation}
\label{sec:harness}

\pp{Paper input.}
All agents receive the PDF, figure crops, and identical line-numbered
text extracted by MOCR with Dots-MOCR~\cite{rednote2026dotsmocr}, reducing reliance on
each agent's ability to read mathematical notation in PDFs. Dots-MOCR scores $83.9 \pm
0.9$ on
olmOCR-Bench~\cite{poznanski2025olmocr}. Shared text reduces reading costs and
supplies stable evidence anchors; document understanding can still vary. Fifteen
invalid claim occurrences across seven papers
were attributed to OCR errors (\S\,\ref{sec:rq3}).

\pp{Audit session.}
Each audit runs in a separate Codex CLI session~\cite{openai2025codex}
with the paper files, a report template, and, when applicable, the playbook.
The auditor can delegate checks to subagents; the supplied
playbook guides their use. An Apple M4 Mac mini with 16~GB memory
orchestrated sessions; LLM inference used remote APIs.

\section{RQ1: Playbook Development}
\label{sec:design}
\label{sec:rq1}

 Figure~\ref{fig:playbook-development} shows how the playbook evolves.

\input{figures/workflows/playbook-development-figure}

\subsection{Playbook Development Process}
\label{sec:playbook-development}
Development used 540 historical training cases
between July and August 2026. Training cases and \DatasetKF{} are separated
at the vulnerable proof-component level (\S\,\ref{sec:data-bench}).
An LLM Reviser proposed playbook revisions from audit reports and grading
feedback (Figure~\ref{fig:playbook-development}). Authors specified feedback
requirements and resolved grading disputes.

When a reported claim matched a previously documented false-positive
pattern, authors required feedback to explain why the claim was unsupported
and what stronger evidence could justify reconsidering it. The Reviser was
instructed to turn transferable explanations into checks that an auditor
could apply using the paper itself. For example, the final accepted
revision required the Auditor to check suspected flaws against the paper's
security definition and the allowed operations of the relevant algorithm.

Each proposed revision was tested by rerunning audits on selected
development cases and comparing their credit with stored baseline reports.
We summed route credit by severity over the same comparable cases,
comparing Critical first, then Medium and Low to break ties. A proposal
was adopted only with a strict improvement; otherwise the current playbook
was retained (Appendix~\ref{app:playbook-comparison}).

Accepted revisions supplied audit tactics and most reporting requirements.
Authors also used the Answerbook on three fixed validation cases to select
the search procedure in
\S\,\ref{sec:anatomy}, prioritizing Critical-route over Medium credit.
The evaluated playbook combines these revisions, the selected search
procedure incorporated before the third training pass, and the multi-agent
coverage protocol added after training.

\subsection{How the Playbook Guides Audits}
\label{sec:anatomy}

The full playbook in our repository tracks open checks and the evidence each finding needs.

\pp{Audit workflow.}
During playbook development, focusing on a promising proof flaw left other
checks unfinished, including construction tests that had exposed flaws in
earlier audits. We introduced the workflow in
Figure~\ref{fig:playbook-audit-workflow} to resume unfinished checks after
each investigation. (i)~The auditor maps each main claim to the definitions,
assumptions, constructions, and proof steps it relies on, tracing local
defects to the claims they affect. (ii)~It selects the untested dependency whose
failure would matter most and records where to resume. It tests the
construction on concrete inputs and checks the premises of the proof
independently. (iii)~When it suspects a flaw, it saves unfinished checks
and pauses the search to investigate. A concern about the construction
prompts a check of the proof, and vice versa. (iv)~It resumes saved checks
and revisits untested dependencies. It reports once all candidates are
resolved, all material dependencies are checked, and a fresh sweep after
the last discovery yields no new independent candidate.

\input{figures/workflows/audit-workflow-figure}

\pp{Targeted checks.}
Grading feedback exposed missed flaws and unsupported claims. Each example links a
resulting instruction to a training paper targeted by the revision.

\input{playbook-checks-list}

\looseness=-1
The auditor follows the shared text workflow throughout each audit and retrieves
indexed checks as candidates arise. These instructions encode expert judgments about
search priorities, reasoning, and evidence.

\section{Datasets and Grading}
\label{sec:benchmark}

\subsection{Datasets}
\label{sec:data-bench}

\pp{Historical development cases.}
Development and known-flaw evaluation use human-confirmed attack--victim
pairs from Cryptology ePrint Archive records. The victim is the audit
target; the attack supplies initial reference evidence. Splitting by
attack-paper date yields 540 training cases
(1999--2024), 43 validation cases (2025 H1), and 36 test cases (2025 H2).
Training and validation support playbook development. For recurring victim
papers, vulnerable proof components remain disjoint between training and
evaluation.

\phantomsection\label{sec:evaluation-set}
\pp{Known-flaw evaluation.}
\DatasetKF{} contains 42 papers, each audited three times by all five models
with and without the playbook. It combines 20 retained cases from the
historical test split with 22 admitted 2026 ePrint pairs whose attack papers
first appeared after 30~April 2026, GPT-6.1 Sol's reported training cutoff.
The historical attacks predate all evaluated models' cutoffs and may have
been seen during pretraining. Later attack dates do not establish that
the victim papers were unseen; Appendix~\ref{app:extended} compares the
two cohorts.

\phantomsection\label{sec:data-wild}
\pp{In-the-wild evaluation.}
\DatasetITW{} contains 121 papers from the 2026 editions of CRYPTO,
EUROCRYPT, PKC, ASIACRYPT, and TCC. We select papers whose earliest known
public version appeared after 30~April 2026 and whose claims fall within
our audit scope. We had not previously audited these papers. For ASIACRYPT and TCC we
use public preprints when proceedings versions are unavailable. We prefer the full
text among collected versions, which may predate the latest version.
Appendix~\ref{app:corrected-constructions} documents a retained
proceedings-version input included in the evaluation. Three of the 121 papers describe
known attacks from the benchmark. The corpus starts without reference answers.
Acquisition, exclusions and version matching appear in
Appendix~\ref{app:dataset-reference-details}.

\subsection{Answerbooks and Human Alignment}
\label{sec:alignment}

\pp{Initial references.}
An Answerbook records adopted reference findings, source evidence, and the
reasoning required for credit. A \emph{reference route} is an adopted scoring
unit for one paper. Its rubric may combine several observations or
alternative proof paths, so route counts represent scoring units rather
than a count of independent security failures. For historical cases, an
LLM uses the attack and victim papers to draft reference answers and
rubrics. One author used LLM-assisted comparisons with the published
attacks to screen the Answerbook entries for all training and validation
cases and all 42 cases in the final known-flaw evaluation.
This pass checked agreement with the documented findings;
detailed technical review addressed selected Critical candidates.

\pp{Case-by-case alignment.}
For each paper, an LLM compares audit findings with the source text,
existing reference answers, and prior human judgments. It applies recorded
review criteria and identifies new candidate flaws or unresolved questions
about validity and credit. Authors investigate selected Critical candidates,
check proof steps and counterexamples, and resolve disputes about admitting
new routes, merging overlapping findings, and setting credit requirements
when existing criteria do not settle them. Unresolved technical disputes
are discussed with another coauthor. We record decisions, supporting evidence,
and conditions for reconsideration in the Answerbook, so subsequent reviews
apply settled judgments and focus on unresolved questions.
Both $+P$ and $-P$ reports contribute candidate findings to
each paper's Answerbook: all five models on \DatasetKF{} and GPT-6.1 Sol
on \DatasetITW{}.
For \DatasetITW{}, reports supply the initial references.

We are conducting quality assurance through random sampling of four
components: \DatasetKF{} Answerbooks, new \DatasetKF{} findings beyond
those documented in the corresponding attack papers, \DatasetITW{}
Answerbooks, and \DatasetITW{} audit findings. We are also engaging external
reviewers to validate Critical findings in samples of \DatasetKF{} papers
and \DatasetITW{} papers. This human validation remains in progress.

\pp{Human effort and adopted references.}
\looseness=-1
We spent about 220 person-hours preparing references, screening candidates, and
reviewing selected cases across development and evaluation
(Appendix~\ref{app:dataset-reference-details}; \S\,\ref{sec:production-human-review}
gives the \DatasetITW{} reviews). The adopted \DatasetKF{}
Answerbook contains 212 routes (81 Critical, 107 Medium, 24 Low);
\DatasetITW{} contains 295 (98 Critical, 133 Medium, 64 Low).

\subsection{Grading}
\label{sec:grading}

\pp{Report grading.}
Grading combines LLM scoring and delegated LLM review with author
adjudication of selected cases. For the $+P$ and
$-P$ reports on \DatasetKF{} and \DatasetITW{}, we use \ModelSol at
\texttt{xhigh} reasoning effort in the Codex CLI. The grader applies
adopted Answerbook criteria and prior human decisions. It flags new findings
or disputed credit requirements not settled by these criteria. Authors
investigate selected Critical candidates and disputes through the review
process in \S\,\ref{sec:alignment}; unresolved credit decisions remain pending.
Decisions and their rationale are recorded
in the Answerbook; affected $+P$ and $-P$ reports are graded or
regraded under the same updated criteria.

\pp{Rubric credit.}
The grader matches findings to reference routes, merges duplicate matches,
and credits the reasoning and evidence supplied by each report. Rubrics
weight proof obligations and specify prerequisites: an unmet prerequisite
blocks credit for dependent obligations. Partial credit captures incomplete
reasoning; alternative proof paths are scored separately, and the highest
normalized path score counts. Each route uses its recorded severity.
Claim validity is recorded separately from route credit: a supported claim
may receive no independent credit, while partial credit does not certify
the entire claim. We report claim-validity outcomes in
Appendix~\ref{app:rq-results-details} and \S\,\ref{sec:rq3}.
Appendix~\ref{app:rubric} gives a worked example, and
Appendix~\ref{app:dataset-reference-details} provides the rubric specification.
\S\,\ref{sec:study} defines how route scores are aggregated into coverage,
overall audit scores, and repeat-detection measures.

\section{Study Design}
\label{sec:study}

\input{figures/workflows/evaluation-workflow-figure}

\pp{Conditions and models.}
On \DatasetKF{} and \DatasetITW{}, we compare auditing without
the playbook ($-P$) and with the human-aligned playbook ($+P$).
Within each model, both conditions use the same prompt template and allow subagents.
This compares the complete playbook procedure, including its
runtime protocol, with the computation and continuation or repair
instructions used in each condition.
It does not isolate domain-specific tactics from generic persistence,
delegation, or additional checking.
On \DatasetKF{}, we evaluate GPT-5.6 Sol, GPT-6 Sol, GPT-6.1 Sol, GLM-5.3-Flash,
and MiniMax-M3.1-Flash, running each model three times per paper with the
playbook and three times without it. The three GPT models use \texttt{xhigh} reasoning
effort; GLM-5.3-Flash and MiniMax-M3.1-Flash use \texttt{high}.
Effort labels are not calibrated across models or providers.
Section~\ref{sec:benchmark} describes the datasets.  On \DatasetITW{}, we use
GPT-6.1 Sol at \texttt{xhigh} effort on 121 papers, with three rollouts
per paper and condition.  The five-model comparison on \DatasetKF{} uses Codex CLI~0.147;
the three GPT configurations also share system, developer, and task
instructions. The \DatasetITW{} runs use Codex CLI~0.166,
with web search and shell network access disabled in all runs.
We grade existing reports against the adopted Answerbook, retaining their
recorded model, condition and execution settings.

\pp{Metrics.}
For the $+P$ and $-P$ comparisons on \DatasetKF{}
and \DatasetITW{}, we measure reference-route coverage and report credit.
For model $m$ and condition $a\in\{+,-\}$,
$s_{p,r,j}$ denotes route $j$'s score in paper $p$, round $r$.

\emph{Set-based metrics} count distinct reference
routes with positive rubric credit, whether partial or full. For the stated
severity and evaluable papers, the credited-route set in round $r$ is
$\mathcal C_m^{(a,r)}=\{(p,j):s_{p,r,j}>0\}$.
Each paper--route pair $(p,j)$ counts once. Coverage is the set size, or as a
fraction, the set size divided by the fixed route count for that severity and cohort.
Set differences give arm-specific routes; the intersection gives routes credited in
both (\hyperlink{finding:2}{Finding~2}).

Union coverage counts each route credited in the selected
runs once. Recurrence counts the runs crediting each route; we report routes credited
in at least one, two, or all three runs.
We report these measures primarily for Critical routes
(Figures~\ref{fig:metric-repeat-union} and~\ref{fig:production-critical}).

For percentile $x\in[0,100]$, let
$v_{p,j,(1)}\le\cdots\le v_{p,j,(n)}$ be the pooled observed scores for
reference route $(p,j)$, including zeros.
We pool 30 scores per route in \DatasetKF{} (five models, two conditions, three rounds) and six in \DatasetITW{} (one model, two conditions, three rounds).
The nearest-rank threshold is
$q_{p,j}(x)=v_{p,j,(\max\{1,\lceil nx/100\rceil\})}$.
Within the fixed reference-route set $\mathcal R$ for the stated severity
and evaluable papers, percentile coverage counts routes satisfying both
$s_{p,r,j}>0$ and $s_{p,r,j}\ge q_{p,j}(x)$; its fraction divides this
count by $|\mathcal R|$.
The positive-credit condition excludes zero scores, while $\ge$ retains ties
at the threshold. Thus p0 recovers positive-credit coverage, and p50 uses the
route's nearest-rank median observed score.
If the median is zero, p50 and p0 coincide for that route
(\hyperlink{finding:2}{Findings~2} and~\hyperlink{finding:4}{4};
Figure~\ref{fig:metric-route-percentiles}).

As an absolute sensitivity endpoint, we also count routes with
full rubric credit, $s_{p,r,j}=1$. This requires all available credit in at
least one declared scoring path within a single report; evidence is not
combined across reports
(Table~\ref{tab:reviewer-full-credit-summary}).

\emph{Score-based metrics} aggregate the route credits defined
in \S\,\ref{sec:grading}.
Distinct route scores are summed within each severity bucket,
yielding a $[\mathrm{Critical},\mathrm{Medium},\mathrm{Low}]$ vector per
paper and rollout for the training score. For evaluation, the overall audit
score divides the total credit across papers by the number of reference
routes in the stated severity category and evaluable set. It lies in $[0,1]$
and reflects both coverage and depth of justification. All reference routes,
including those with zero credit, enter the denominator.
Every reference route receives equal weight in the overall
audit score. We average the three round-level scores, weighting
models equally when reporting a cross-model mean.
For a fixed reference set and at least one covered route, each round's
score is coverage fraction times mean credit on covered routes.
The \emph{shared-positive comparison} pairs observations
by model, paper, route, and recorded round. We average scores by condition over pairs
credited in both $+P$ and $-P$
(\hyperlink{finding:3}{Finding~3}; Figure~\ref{fig:metric-shared-hit-pairs}).
Missing or unresolved grades are reported separately from zero scores;
paired comparisons use common evaluable papers and routes.

Neither coverage nor score subtracts a separate false-positive
penalty. Invalid claims earn no credit for their conclusions.
Supported evidence retains its independently
awarded credit within the same report.
Incorrect reasoning earns no credit for affected obligations, and unmet
prerequisites block dependent credit.
Both metrics use the adopted reference.

For model $m$, condition $a$, and recorded round
$r$, write $S_m^{(a,r)}$ for the overall audit score at the stated severity.
Within each model,
\emph{$+P$-only} and \emph{$-P$-only} routes are
$\mathcal C_m^{(+,r)}\setminus\mathcal C_m^{(-,r)}$ and the reverse
difference. We distinguish \emph{Score gain},
$\Delta S_m^{(r)}=S_m^{(+,r)}-S_m^{(-,r)}$, from
\emph{Net coverage gain}, $\Delta N_m^{(r)}=|\mathcal C_m^{(+,r)}|-
|\mathcal C_m^{(-,r)}|$, which equals the $+P$-only count minus the
$-P$-only count
(\hyperlink{finding:2}{Finding~2}).

For single-rollout performance, we aggregate all papers by rollout index
for both metric families. We report the three totals and their mean, minimum, and
maximum.
To combine $k$ rollouts, we evaluate all
$\binom{3}{k}$ subsets of recorded rounds, using the same subset for every
paper. Within the stated severity, we count each paper's distinct routes
credited in any selected round, then sum across papers. We report the mean,
minimum, and maximum of these totals for $k=1,2,3$
(Figures~\ref{fig:metric-repeat-union} and~\ref{fig:production-critical}).

\pp{Cross-configuration coverage.}
For model--condition configurations $i=(m,a)$ and $j$,
write $\mathcal C_i^{(r)}=\mathcal C_m^{(a,r)}$. At each recorded rollout
index, the directed coverage difference is
$D_{ij,r}=|\mathcal C_i^{(r)}\setminus\mathcal C_j^{(r)}|$. The matrix entry
is $\overline D_{ij}=\frac{1}{3}\sum_{r=1}^{3}D_{ij,r}$. It counts routes
covered by the row configuration but not the column configuration. The two
directions generally differ (Figure~\ref{fig:metric-directed}).
Rollout indices pair runs descriptively, not by random seed.
We report missingness; pairs use common evaluable papers and routes.

\pp{Combined overall audit score.}
On \DatasetKF{}, we compare how to allocate three playbook
audits of each paper across the five models, using combined route scores
and inference cost. An allocation specifies the number of audits assigned
to each model: $n_1,\ldots,n_5\ge0$ are integers with
$\sum_{i=1}^{5}n_i=3$, giving $\binom{3+5-1}{3}=35$ allocations.
For each Critical reference route, we select
the highest score among its paper's three reports, then average over all
81 reference routes, including those with zero credit. This retrospective
selection does not combine evidence across reports.

For each allocation, we average combined
scores and summed inference costs over all available combinations of
recorded runs, with equal weight. Each combination uses the same runs across papers
and distinct rounds for repeated models. Total cost divided by 42 gives three-audit
cost per paper
(\hyperlink{finding:5}{Finding~5}; Table~\ref{tab:three-run-cost-score}).

\pp{Repeat detection of previously credited routes.}
For each model, each of the three $-P$ rounds
in turn supplies a checklist of Critical reference routes with positive
credit (p0). We set aside the corresponding $+P$ and $-P$ rounds.
Separately for $+P$ and $-P$, we divide the number of checklist routes
meeting the chosen threshold (p0 or p50) in both remaining rounds by that
checklist's size, then average the three fractions equally. All checklists
are nonempty. The checklist always uses p0; p50 changes only the evaluation
threshold. We retain the p50 thresholds defined above, whose score pool
includes the evaluation runs. These descriptive estimates summarize three
overlapping splits of the same recorded runs
(\hyperlink{finding:4}{Finding~4}; Table~\ref{tab:baseline-repeatability}).

We also report coverage's sample standard
deviation, mean pairwise Jaccard similarity, and the three-run intersection
divided by the union.
These ratios use different credited-route sets across
conditions. Intermittently credited additional routes can lower overlap even
when previously credited routes recur more reliably
(\S\S\,\ref{sec:rq2} and~\ref{sec:rq3}).

\emph{Claim validity} is reported separately as supported,
confirmed invalid, unverified, or pending.
Outcome proportions use reviewed claim
occurrences, including repeats across reports. Legacy finding labels and
reviewed claim labels classify different units
(Appendix~\ref{app:rq-results-details} and \S\,\ref{sec:rq3}).
For \DatasetKF{} cost--coverage, we price recorded token usage at a fixed historical
API price card, including accepted Auditor roots, children, continuations,
and repairs; grading and separate earlier failed attempts are excluded.
\S\,\ref{sec:limitations} describes incomplete usage records and the
estimation policy
(\hyperlink{finding:5}{Finding~5};
Figure~\ref{fig:metric-cost-coverage}).

\pp{Statistics.}
Table~\ref{tab:critical-overview} and Table~\ref{tab:production-performance}
report means and gains computed before rounding; gain ranges are observed
round minima and maxima, not confidence intervals.
$+P$ and $-P$ are compared within each model and paired by paper.
In \S\,\ref{sec:rq3}, wins, ties, and losses
denote higher, equal, and lower three-round mean Critical score totals
for $+P$ than $-P$.

\section{RQ2: Effectiveness}
\label{sec:rq2}

Each of five models audits all 42 \DatasetKF{} papers
three times with and three times without the playbook, producing
$42\times5\times3\times2=1{,}260$ audit reports. We ask whether the
playbook helps models identify Critical flaws
(cf.\ \S\,\ref{sec:taxonomy-severity}) and support their findings with evidence.
We highlight the key insights below, labeled \ref{insight:model-rankings},
\ref{insight:model-gap}, and so on.

\subsection{\texorpdfstring{Overall Performance (\protect\hyperlink{finding:1}{Finding~1})}{Overall Performance (Finding 1)}}
\label{sec:rq2-performance}

The playbook increases the Critical overall audit score
in all 15 model--round comparisons (Table~\ref{tab:critical-overview}).
Weighting the five models equally, mean score rises from 0.209 to 0.390
($+86.9\%$); model-level mean gains range from 0.121 to 0.273.
Under the initial 58-route Critical reference, score gains
were 0.211 for GPT-5.6 Sol and GPT-6 Sol, 0.038 for GPT-6.1 Sol, 0.127 for
GLM-5.3-Flash, and 0.247 for MiniMax-M3.1-Flash
(Table~\ref{tab:appendix-sensitivity-summary}). This reference already
incorporated expert judgments.
Each model earns more Critical credit with the playbook
on 19--29 of the 35 papers with known Critical flaws.
Mean score and coverage also increase in Medium and Low, with full results
in the artifact. These comparisons measure the complete procedure at observed, unequal
compute.
Grading outcomes for individual claims appear in
Appendix~\ref{app:rq-results-details}.

\input{tables/benchmark-critical-results.tex}

\begin{figure}[tb]
  \centering
  \includegraphics[width=\columnwidth]{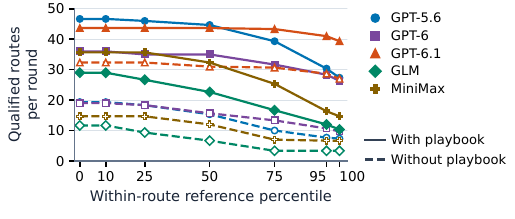}
  \caption{Mean qualified Critical routes per round on \DatasetKF{}.
  Thresholds use 30 scores per route (five models, two conditions, three rounds).
  Lines join the evaluated percentiles.}
  \label{fig:metric-route-percentiles}
\end{figure}

\begin{figure*}[tb]
  \centering
  \includegraphics[width=\textwidth]{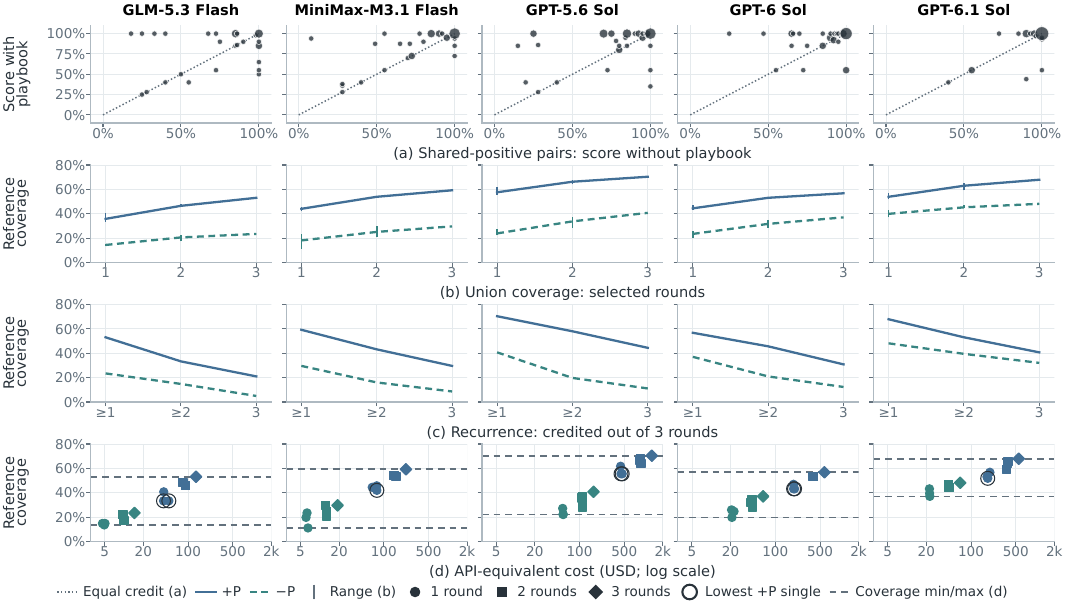}
  \caption{Reviewed Critical results on \DatasetKF{}
    (81 reference routes). (a) Paired scores with positive credit in both arms for the
    same paper, route and round; marker area reflects frequency. (b) Mean and min--max
    union coverage over $3,3,1$ subsets of $k=1,2,3$ rounds. (c) Coverage in at least
    one, two or all three rounds. (d) Coverage versus summed API-equivalent cost (log
    scale) for all seven nonempty subsets per arm. Rings mark the lowest-coverage $+P$
    singles; dashed horizontal lines mark both arms' extrema. In (b--d), blue is $+P$,
    teal $-P$. Coverage counts any positive credit out of 81 routes. Subsets are fixed
    across papers; ranges are not confidence intervals.}
  \label{fig:metric-shared-hit-pairs}
  \label{fig:metric-repeat-union}
  \label{fig:metric-persistence}
  \label{fig:metric-cost-coverage}
\end{figure*}

\insight{Playbook gains do not follow baseline rankings.}\label{insight:model-rankings}
GPT-5.6 Sol has the largest absolute score gain ($+0.273$),
despite starting below GPT-6 and GPT-6.1. With the playbook,
it scores above GPT-6 (0.469 versus 0.400) but below GPT-6.1
(0.506), while leading coverage at 57.6\%
(\S\,\ref{sec:rq2-coverage}).

\insight{The playbook narrows gaps to the strongest model.}\label{insight:model-gap}
With the playbook, all four other models move closer to
GPT-6.1 in overall score. GPT-6.1 gains 31.5\%, yet its lead
over GPT-5.6 shrinks from 0.188 to 0.036.

To understand these score gains, we examine how many Critical
routes the auditors find (\S\,\ref{sec:rq2-coverage}) and how fully they
justify the same findings with and without the playbook (\S\,\ref{sec:rq2-depth}).
We then measure how often findings recur~(\S\,\ref{sec:rq2-repeat}).

\subsection{\texorpdfstring{Coverage (\protect\hyperlink{finding:2}{Finding~2})}{Coverage (Finding 2)}}
\label{sec:rq2-coverage}

All five models found more Critical flaws with the playbook in
each of the three rounds (Table~\ref{tab:critical-overview}). GPT-6.1 Sol went from
32.3 to 43.7 of the 81 Critical routes per round on average. Its full-credit
detections rose from 26.7 to 38.0 (Table~\ref{tab:reviewer-full-credit-summary}).
GPT-5.6 Sol covered the most routes with the playbook (46.7 per round).
At the stricter p95 threshold, GPT-6.1 led with 41.0 routes to GPT-5.6's
30.3 (Figure~\ref{fig:metric-route-percentiles}).

\pp{Invalid claims and missed routes.}
For false positives, we report the share of claims in
findings labeled Critical that the grader marked invalid, counting repeats.
For false negatives, we report the share of the 81 Critical reference routes
that received zero credit in all three runs of a condition.
With the playbook, all five models missed fewer routes. Invalid claims
rose from 103 to 170, and their share fell from 11.99\% to 8.58\%
(Table~\ref{tab:test-critical-errors}).\looseness=-1

\begin{table}[tb]
  \centering\footnotesize
  \caption{Invalid claims and missed Critical routes on the 42-paper
    known-flaw benchmark.}
  \label{tab:test-critical-errors}
  \setlength{\tabcolsep}{2pt}
  \begin{tabular*}{\columnwidth}{@{\extracolsep{\fill}}lrrrr@{}}
    \toprule
    & \multicolumn{2}{c}{Invalid Critical claims}
    & \multicolumn{2}{c}{Missed routes (of 81)} \\
    \cmidrule(lr){2-3}\cmidrule(l){4-5}
    Model & $-P$ & $+P$ & $-P$ & $+P$ \\
    \midrule
    5.6 Sol & 10/173 (5.78\%) & 17/448 (3.79\%) & 48 (59.3\%) & 24 (29.6\%) \\
    6 Sol & 8/135 (5.93\%) & 8/278 (2.88\%) & 51 (63.0\%) & 35 (43.2\%) \\
    6.1 Sol & 1/125 (0.80\%) & 0/223 (0.00\%) & 42 (51.9\%) & 26 (32.1\%) \\
    GLM & 40/251 (15.94\%) & 69/645 (10.70\%) & 62 (76.5\%) & 38 (46.9\%) \\
    MiniMax & 44/175 (25.14\%) & 76/387 (19.64\%) & 57 (70.4\%) & 33 (40.7\%) \\
    \bottomrule
  \end{tabular*}
\end{table}

We paired reports by paper, model, and round.
Table~\ref{tab:coverage-misses} shows what the other audit checked
when only one report earned any credit for a Critical flaw.

\begin{table}[tb]
  \centering
  \caption{Failure modes for missed Critical flaws.}
  \label{tab:coverage-misses}
  \small
  \setlength{\tabcolsep}{3pt}
  \begin{tabularx}{\columnwidth}{@{}Yrr@{}}
    \toprule
    Failure mode &
      $+P$ only &
      $-P$ only \\
    \midrule
    Pursued a related question & \textbf{165 (53.9\%)} & \textbf{17 (70.8\%)} \\
    Raised but left open & 3 (1.0\%) & 1 (4.2\%) \\
    Drew a wrong conclusion & 9 (2.9\%) & 1 (4.2\%) \\
    Explicitly deferred the issue & 0 (0\%) & 3 (12.5\%) \\
    Unclear from trace & 129 (42.2\%) & 2 (8.3\%) \\
    \midrule
    Total & 306 & 24 \\
    \bottomrule
  \end{tabularx}
\end{table}

\Needspace{3\baselineskip}
\insight{The playbook helps models find flaws in places they already check.}\label{insight:coverage-same-step}
In 165 of the 306 pairs where only $+P$ found the flaw, the
$-P$ audit checked a different question at the same step. On P035,
GPT-6.1 checked whether the verifier's computation was internally
consistent. With the playbook it also asked whether that computation
used the public input for this particular run. Even a consistent
computation can use the wrong input. All three $+P$ reports earned
full credit for this flaw, and all three $-P$ reports missed it.

\insight{Did the model notice the clue? Sometimes we are not sure.}\label{insight:coverage-unclear}
In another 129 of the 306 $+P$-only pairs, the $-P$ audit's
tools returned the relevant passage. On P030, for example, GPT-6's
search for a symbol brought up the key line. Nothing later in the
trace showed whether the model noticed the issue. Overlooking a clue
and deciding it is too weak to report can leave the same record.

\insight{Sometimes models follow a clue further without the playbook.}\label{insight:coverage-follow-clue}
On P024, GPT-6 Sol's playbook audit flagged a mismatch
between the distribution used in the paper and the one assumed by a
cited result. It left that question open and pursued another step in
the proof. Without the playbook, GPT-6 showed why the cited result
did not apply to the paper's distribution.
In 19 of the 24 $-P$-only pairs, the same model found the flaw in
another $+P$ run.

\insight{Even with the playbook, some problems were still too hard for the models.}\label{insight:critical-misses}
All 30 audits of P012 missed its one known Critical flaw,
as did all 30 audits of P022. On P012, GPT-6.1 found the same proof error
in all three playbook runs. Its checks included verifying that test
messages decrypted correctly. Those tests could pass even when an
observer could distinguish the construction from ideal encryption.
On P022, one GPT-6.1 playbook audit checked the proof assuming the
underlying problem was hard. At the paper's chosen parameters, that
problem was easier than the proof assumed. These runs leave open
whether more specific guidance would have helped or whether the models
would still have struggled even if prompted to check the right premise.

\subsection{\texorpdfstring{Depth of Justification (\protect\hyperlink{finding:3}{Finding~3})}{Depth of Justification (Finding 3)}}
\label{sec:rq2-depth}

The Answerbook's route score (\mbox{0--1}, cf.\ \S\,\ref{sec:grading})
measures how fully a report justifies a finding. Figure~\ref{fig:metric-shared-hit-pairs}a
pairs audits by model, paper, and round to compare scores on Critical
routes credited with and without the playbook.

\insight{The playbook helps lower-scoring models fully justify flaws they already find.}\label{insight:depth-shared}
For each of the four models that scored below GPT-6.1 Sol
without the playbook (\S\,\ref{sec:rq2-performance}), more matched
findings moved from partial to full credit than the reverse. Of the
104 matched detections these models only partly justified without the
playbook, 51 earned full credit with it. Fourteen previously complete
arguments fell to partial credit.
Mean credit on matched findings rose by 5.3--10.7\%.

\insight{GPT-6.1 Sol gains by substantiating flaws it misses without the playbook.}\label{insight:depth-additional}
Mean credit on its 90 matched detections stayed near the
ceiling (0.972 without the playbook, 0.970 with it). Across three rounds,
41 detections appeared only with the playbook and seven only without it.
Of those 41 additional detections, 32 earned full credit.

Across all five models, mean credit per detection was lower
with the playbook when we included everything each audit found.
The additional detections earned less credit on average than the matched
detections, changing the mix of findings in that average
(Appendix~\ref{app:rq-results-details}).

\subsection{\texorpdfstring{Repeat Detection (\protect\hyperlink{finding:4}{Finding~4})}{Repeat Detection (Finding 4)}}
\label{sec:rq2-repeat}

If a model finds a flaw once, will it find it again?
For each model, we make a list from one run without the playbook.
A Critical route on that list counts as a repeat if the model finds it
in \emph{both} other rounds. We compare the repeat rates with and
without the playbook (\S\,\ref{sec:study}).

\insight{The playbook helps auditors find the same flaws again.}\label{insight:repeat-overlap}
Without the playbook, GPT-5.6 Sol found 47\% of its listed
routes in both repeat audits. With the playbook, it found 88\%.
The playbook helped all five models, whichever run we used to make
the list. The gains also held at the higher evidence threshold in
Table~\ref{tab:baseline-repeatability}.

\input{tables/baseline-repeatability.tex}

\insight{One run with the playbook found more flaws than three runs without it.}\label{insight:repeat-union}
Across three runs without the playbook, GPT-5.6 Sol found
33 distinct Critical routes in total. Its lowest count in a single run
with the playbook was 45. The same held for each of the five models
(Figure~\ref{fig:metric-repeat-union}b).
Depending on the model, one playbook run cost on average
$2.7$--$3.9\times$ as much as three baseline runs. Whether more baseline runs
would close the coverage gap at the same inference cost remains open
(\S\,\ref{sec:limitations}).

\subsection{\texorpdfstring{Cost and Model Allocation (\protect\hyperlink{finding:5}{Finding~5})}{Cost and Model Allocation (Finding 5)}}
\label{sec:rq2-cost}

\begin{figure*}[t]
  \centering
  \includegraphics[width=\textwidth]{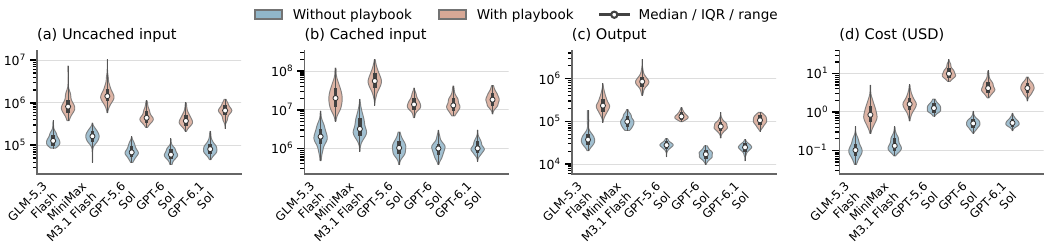}
  \caption{\DatasetKF{} token use and API-equivalent USD per audit,
    including children and repairs: 42 papers, three rounds, 126 reports per
    model and arm. Log-scaled axes show positive values; violin widths encode
    log-space densities. Dots mark medians, thick lines interquartile ranges,
    and thin lines full ranges. Input separates cached and uncached tokens;
    output includes reasoning. Repeated reports are descriptive.}
  \label{fig:test-token-usage}
\end{figure*}

What does a repeat audit add, and when does switching models
help? We count distinct Critical routes across runs. For three playbook
audits (\S\,\ref{sec:study}), we sum estimated inference costs and
average each route's best single-report score.
Figure~\ref{fig:test-token-usage} shows the cost per audit.

\insight{Rerunning the same model recovers missed flaws, with diminishing returns.}\label{insight:cost-repeat}
With the playbook, one GPT-6.1 Sol run covered 43.7 routes
on average. Two runs together covered 51 on average, and all three
covered 55 (Figure~\ref{fig:metric-repeat-union}b). Each of the five
models added fewer routes on the third run than on the second when
averaged over run order. Across models, three playbook runs covered
$1.41\times$--$2.26\times$ as many routes as three runs without the
playbook, at $8.1\times$--$11.7\times$ the estimated inference cost
(Figure~\ref{fig:metric-cost-coverage}d).

\begin{figure}[tb]
  \centering
  \includegraphics[width=\columnwidth]{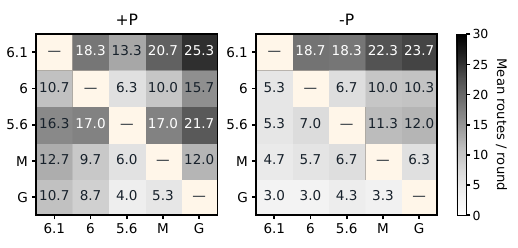}
  \caption{Pairwise Critical-route complementarity on \DatasetKF{}
    under reviewed grading. Each other model credits 10.7--16.3 routes per
    round absent from the highest-scoring GPT-6.1 Sol with the playbook
    ($+P$, left), versus 3.0--5.3 without it ($-P$, right).
    Cells show three-round mean routes credited to the row but not the
    column (p0, including partial credit); darker means more routes on a shared scale.
    6.1/6/5.6: GPT Sol versions; M: MiniMax-M3.1-Flash;
    G: GLM-5.3-Flash.}
  \label{fig:metric-directed}
\end{figure}

\insight{Some flaws recur for one model and go undetected by another.}\label{insight:model-coverage}
Figure~\ref{fig:metric-directed} compares the routes each
model found in the same round. Five Critical routes appeared in all
three GLM playbook audits and in none of the GPT-6.1 Sol audits.
Eleven other routes showed the reverse pattern. We also observed
differences in both directions when comparing GPT-6.1 Sol with each
of the remaining three models.

\insight{A cheaper model can add more evidence on the third run.}\label{insight:cost-third}
We hold two GPT-6.1 Sol runs fixed and vary the model for
the third. Averaged over the recorded combinations, another GPT-6.1 Sol
run added four routes beyond the first two. Each alternative added
more, from 8.3 routes for GLM to 11.6 for GPT-5.6 Sol. GLM ranked
last in single-run score (Table~\ref{tab:critical-overview}), yet replacing one
GPT-6.1 Sol run with GLM raised the mean combined score from 0.639
to 0.652 and lowered estimated cost per paper from \$13.25 to \$9.89.
Table~\ref{tab:three-run-cost-score} lists the lower-cost combinations
retaining GPT-6.1 Sol. The artifact reports all 35 allocations.

\input{tables/three-run-cost-score.tex}

\subsection{Component Ablation}
\label{sec:rq2-ablation}

The full playbook improves audit performance
(\S\S\,\ref{sec:rq2-performance} and~\ref{sec:rq2-repeat}).
How much of that gain do we see with its search and verification protocol alone?
We test three conditions.

\begin{enumerate}[label=\arabic*),leftmargin=*,nosep]
  \item \textbf{No playbook.} The auditor chooses what to inspect, how
  to verify candidates, and whether to delegate.
  \item \textbf{Protocol only.} The protocol requires the lead to map
  the main claims (Figure~\ref{fig:playbook-audit-workflow}, stage~1)
  and assign at least two agents to complementary searches (stage~2).
  The harness requires a record showing that a third agent verified candidates (stage~3),
  searchers resumed unfinished work (dashed return arrow), and the
  verifier made a fresh sweep after the last discovery (stage~4).
  \item \textbf{Full playbook.} The playbook adds a rule for stage~2
  that prioritizes the untested dependency whose failure would most affect
  a main claim. It also supplies targeted checks for stage~3, such as
  comparing the success probability over repeated attempts with the
  theorem's bound (\S\,\ref{sec:anatomy}).
\end{enumerate}

We select two papers from each of six groups defined by attack
source and Critical route type (Appendix~\ref{app:rq-results-details}).
Each of the twelve papers contributes three reports per condition,
for 108 reports in total. All reports use GPT-6.1 Sol/\texttt{xhigh}
with Codex CLI~0.166.
We grade all reports against the same frozen Answerbook
(cf.\ \S\,\ref{sec:alignment}).
Table~\ref{tab:component-ablation} weights papers equally. For single
runs we report the mean and sample standard deviation of the three
round averages. To combine runs, we take each route's best single-report
score (\S\,\ref{sec:study}).

\input{tables/component-ablation}

\insight{The protocol helped GPT-6.1 find more flaws and find them again.}\label{insight:ablation-coverage}
With the search and verification protocol
(Figure~\ref{fig:playbook-audit-workflow}), mean coverage per run rose
from 34.5\% to 51.6\% and mean audit score from 0.322 to 0.489.
Coverage increased on six papers and held steady on the other six.
Across three runs, the protocol found all 13 Critical routes detected
without the playbook and six more. All four routes detected only once
or twice without the playbook appeared more often with the protocol.

\Needspace{3\baselineskip}
\insight{The full playbook broadened coverage across runs, but fewer flaws recurred in all three.}\label{insight:ablation-full-playbook}
Adding the priority rule and targeted checks left the mean
score per run nearly unchanged (0.489 to 0.485). Coverage varied more
from run to run (SD 1.6 to 12.1 percentage points). Across three runs,
coverage rose from 63.2\% to 74.3\% and the combined score from 0.616
to 0.665. The full playbook found all 19 routes detected by the
protocol alone and two more, one each on P027 and P042. Each extra
route appeared only once. Of the 19 shared routes, the protocol alone
found 14 in all three runs and the full playbook found 12.
Whether targeted guidance can extend GPT-6.1's coverage while preserving
repeat detection remains open.

\section{RQ3: In-the-Wild Audit Performance}
\label{sec:rq3}

RQ3 tests the playbook on papers selected without requiring a known flaw:
726 GPT-6.1 Sol reports on 121 \DatasetITW{} papers across two conditions and three
rounds.
The Answerbook draws its reference routes from reports in both conditions
(\S\,\ref{sec:alignment}). GPT-6.1 Sol grades both conditions against that
reference. Coverage is measured within this pool.

\subsection{Performance Beyond Known-Flaw Papers}

\input{tables/production-performance.tex}

Critical overall score rises from 0.433 to 0.652, and mean
positive-credit coverage rises from 46.6\% to 71.1\% of the 98 Critical
reference routes (Table~\ref{tab:production-performance}). Compared with the baseline, the three-round mean total Critical credit per paper was higher with the playbook on 33 papers, equal on 83, and lower on 5. Of the 83 ties, 64 papers had no Critical reference route and therefore scored zero in both conditions.

Across the three paired rounds, a mean of 30.67 Critical routes received
credit only with the playbook and 6.67 only without it
(Figure~\ref{fig:production-critical}). Thus, even with higher overall coverage,
playbook reports missed some routes credited to baseline reports in the same round.
For each Critical route in a paper, p50 requires a positive score at least as high as
the third-lowest of its six scores across both conditions, including zeros. Partial
evidence can meet this threshold. At p50, the playbook covered 26, 23, and 26 more
routes across paired rounds.
Full credit (score 1) requires one complete path in one report. The mean full-credit
route count rose from 38.33 to 56.00 per round with the playbook
(Table~\ref{tab:reviewer-full-credit-summary}, Appendix~\ref{app:extended}; detailed
counts and ranges in Appendix~\ref{app:rq-results-details}).

\pp{One example: a correctness failure in distributed decryption.}
Reports from both audit conditions identified a flaw in distributed decryption in the CRYPTO 2026 proceedings version of Afshar and Goyal's paper on multi-key fully homomorphic encryption~\cite{DBLP:conf/crypto/AfsharG26}. Each party computes a share of the decryption result. The proof assumes that the parties derive their shares from the same ciphertext tree, but the parties hardcode independently sampled smudging values into their evaluation circuits and can build different trees. Offsets that cancel for contributions from the same parent ciphertext need not cancel across parents. Four of the six reports gave counterexamples in which each complete parent ciphertext decrypts correctly, yet combining shares from different parents can give an incorrect result. The counterexamples refute the generic construction's correctness guarantee even after correction of a separate argument-order typo. In the associated second version on ePrint~\cite{afshar2026succinct}, the authors acknowledge the error and describe how to derive a common ciphertext tree using deterministic FHE evaluation, correlation-intractable hashing, and a trusted setup. This example concerns the proceedings construction. Discovery priority and the correctness of the revised construction lie outside its scope. Appendix~\ref{app:corrected-constructions} explains the counterexample, traces the correction, and lists all six report scores by round.

\subsection{Coverage Across Repeated Audits}
\label{sec:rq3-runtime}

\begin{figure}[tb]
  \centering
  \includegraphics[width=\columnwidth]{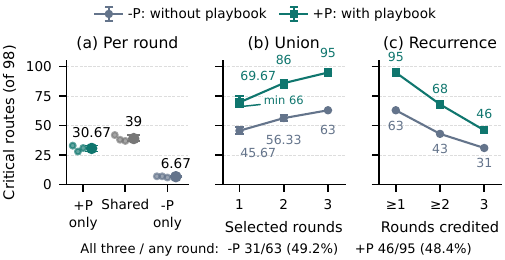}
  \caption{\DatasetITW{} (GPT-6.1 Sol, 98 Critical routes). (a)
  Routes credited to one or both arms. Marks show paired rounds (small) and
  means (large). Whiskers span the range. (b) Union coverage (mean and range)
  over all $k$-round subsets fixed across papers. (c) Recurrence. Fractions give
  the share of each arm's detected routes credited in all three rounds.}
  \label{fig:production-critical}
\end{figure}

Increasing the number of baseline audit rollouts from one to three did not close the coverage gap with the playbook (Figure~\ref{fig:production-critical}b). We compared coverage from one, two, and three audit rollouts per paper in each condition without matching inference costs. Across three rounds, playbook reports covered 95 of the 98 Critical reference routes at least once. The three baseline runs together covered 63 of 98, below the single-run playbook minimum of 66. \newline Per-round~coverage was 49, 45, and 43 without the playbook and 75, 66, and 68 with it.

Our earlier three-rollout intersection analysis examined which routes the
same configuration recovered in every rollout
(Figure~\ref{fig:metric-shared-hit-pairs}(c)). On \DatasetITW{}, 31 of the 63 routes
credited at least once without the playbook were credited in all three rollouts
(49.2\%), compared with 46 of 95 routes with the playbook (48.4\%;
Figure~\ref{fig:production-critical}(c)). The playbook covered more routes, with a
similar fraction recurring within each configuration's own discoveries.

Label the three baseline and playbook rounds BR1--BR3 and PR1--PR3. To
compare repeat detection on the same routes, take the 49 routes credited at p0 in BR1.
At each threshold (p0 or p50), compare the fraction qualifying in both BR2 and BR3
with the fraction qualifying in both PR2 and PR3. For example, 31 of these 49 routes
qualify in both BR2 and BR3, and 36 qualify in both PR2 and PR3.

We repeat with BR2 and BR3 as checklists, excluding the checklist round
from both configurations. Across the three equally weighted checklists, repeat
detection increases from 68.1\% without the playbook to 78.2\% with it at both p0 and
p50 (Table~\ref{tab:baseline-repeatability}). Appendix~\ref{app:rq-results-details}
details overlap and variability.

\subsection{Grading and Human-Review Coverage}

\input{tables/production-claims.tex}

Table~\ref{tab:production-claims} summarizes LLM grading and
delegated LLM review. The 726 reports contained 2,600 finding occurrences
and 2,658 claim occurrences because a finding could contain several claims.
Repeated claims count once per occurrence. Graders assessed validity
separately from credit: 1,490 supported claim occurrences earned no
independent route credit, so zero credit alone cannot identify a false
positive. Six others received final scores of zero while their scientific
validity remained unresolved (Appendix~\ref{app:rq-results-details}).
Eight reports listed checks but no issues; they certify neither complete
audits nor flaw-free papers.
\phantomsection\label{sec:production-human-review}
The grader attributed all 15 invalid judgments on seven
papers to OCR, without independent PDF verification.
Two experts prioritized severe findings. The grader marked claims
for all 16 findings with notes on 13 papers as supported. The notes supported
3 central conclusions and directly corroborated 1 of 6 linked Critical routes
(Appendix~\ref{app:rq-results-details}).

\section{Discussion}
\label{sec:discussion}

\pp{How well does the playbook generalize?}
All five models on the 42-paper test set found more Critical
flaws with the playbook and found them again more often. The four
lower-scoring models also narrowed the gap to GPT-6.1 Sol
(\S\,\ref{sec:rq2}). On 121 papers, GPT-6.1 gained coverage and repeat
detection against the pooled reference (\S\,\ref{sec:rq3}).
These gains extend beyond the development cases under our definition
of the audit task. For example, we distinguish a missing assumption
(Critical) from a misnamed one (Low) in Table~\ref{tab:issue-taxonomy}.
A group with different goals can choose which flaws to look for, how
to group and prioritize them, and what evidence earns credit. It can use those
criteria to grade development audits, select playbook revisions
(\S\,\ref{sec:playbook-development}), and evaluate the resulting
playbook on held-out cases.

\pp{Finding more Critical flaws remains the main challenge.}
On the 42-paper test set, GPT-6.1 Sol missed 26 of the 81
Critical reference routes in all three playbook runs
(\S\,\ref{sec:rq2-cost}). The models already gave strong explanations
for flaws they found both with and without the playbook. On those findings,
mean justification scores with the playbook ranged from 0.81 to 0.97
out of 1 across the five models (\S\,\ref{sec:rq2-depth}).
Future work should test how to detect and justify these missed flaws.

\pp{Do we still need human auditors to find flaws in cryptographic proofs?}
For now, yes. On two test papers, all five models missed a
previously documented Critical flaw in every run, with and without the
playbook (\S\,\ref{sec:rq2-coverage}). We graded the reports against
Answerbooks we developed and refined through expert review. Across the
42 papers, we classified 840 of 8,698 reported claim occurrences as
invalid, counting repeats (Appendix~\ref{app:rq-results-details}).
Experts still have to decide which proof steps deserve closer scrutiny
and check whether a reported error undermines the paper's security claim.
However, the playbook improved GPT-6.1's mean Critical audit score by
0.121, the smallest gain among the five models (Table~\ref{tab:critical-overview}).
Whether the current playbook will help more capable models remains open.
They may need a revised playbook that targets the flaws they still miss,
or they may audit proofs just as effectively without one.

\pp{How can we build larger benchmarks of verified proof flaws?}
A published analysis may identify one flaw in a proof while
others remain undiscovered. Benchmarks built from such analyses tell
us how often auditors recover documented flaws, but little about the
difficulty of finding unknown ones. Preparing reference answers and
reviewing selected findings took about 220 person-hours across development
and evaluation (\S\,\ref{sec:alignment}). An auditor that treats its
own reports as ground truth could reward a false alarm and receive no
feedback on a flaw it repeatedly misses. We therefore lack a reliable
basis for fully automated playbook revision. A priority for future work
is to develop an oracle that examines the proof independently and gives
verifiable reasons for its judgments. By flagging false alarms and missed flaws, such an oracle
could provide reliable feedback on more papers.

\subsection{Responsible Disclosure and Community Validation}
\label{sec:responsible-disclosure}

To support responsible disclosure and broader human validation, we will
maintain a GitHub-hosted index of potential issues\footnote{\url{https://disproofailab.github.io/A0-Artifacts/}}, each corresponding to an
accepted reference route in the adopted \DatasetITW{} Answerbook used in
our evaluation. The full in-the-wild Answerbook will remain private.
The public index will cover every accepted reference route. Its HTML
pages and JSON records will omit the technical details and rubric content
of individual routes. The index will group potential issues by potential
severity (Critical, Medium, or Low), with Critical shown by default.
Potential severity uses the Answerbook's taxonomy-based classification to
estimate impact if the issue is valid. Each entry will have an opaque ID,
its taxonomy category, and separate human-review and disclosure statuses.
Human-review records will distinguish review progress, the reviewer's
assessment of the concern, and the disclosure decision; a decision to
withhold disclosure will include a reason. A recorded human review may leave
the concern's validity unresolved. Disclosure status will record
author contact, pending responses, and author confirmation of the associated
concern as a true or false positive.

For each potential issue, discovery indicators will show whether the
baseline and playbook conditions received positive credit for the
corresponding route in any of the three evaluated rounds.
A SHA-256 hash of the exact
original UTF-8 bytes of each accepted-route JSON object, including its
taxonomy, severity, and rubric, will represent the withheld content.

Our capacity for detailed human verification is limited, and the underlying
concerns may still be false positives. Inclusion records adoption as an
evaluation reference; it does not assert that a paper is flawed. Route
classification and observed audit credit are separate from human validation
and author confirmation.
We encourage authors whose work is listed to contact
\href{mailto:security@disproof.ai}{\texttt{security@disproof.ai}} if they are
interested in using \toolname{} to audit their work. We welcome private
discussion, clarifications, corrections, and supporting evidence. Even
partial responses would help us refine our assessments and update the
review and disclosure records. We appreciate the community's assistance in
validating these concerns. Appendix~\ref{app:dataset-reference-details}
records current disclosure progress and author feedback.

\section{Threats to Validity}
\label{sec:limitations}

\pp{Sampling and pretraining.}
Differences in sampling, reference construction, and flaw mix
(\S\,\ref{sec:data-bench}) prevent us from attributing the larger
relative gain on \DatasetITW{} to paper age. Pretraining may include
historical attacks or papers attacked after a model's cutoff.
\DatasetITW{} uses GPT-6.1 Sol's cutoff and each paper's earliest known
public version. Three runs give a limited estimate of variability.

\pp{Reference construction.}
The six evaluated \DatasetITW{} runs supplied all
98 Critical reference routes (35 $+P$ only, 3 $-P$ only, 60 shared).
Coverage uses this pool, so recall remains unknown. The $+P$ reports
contained 1,609 claim occurrences versus 1,049 for $-P$. On \DatasetKF{},
scores rose under both the initial 58-route and reviewed 81-route
Critical references (Table~\ref{tab:appendix-sensitivity-summary}).

\pp{Grading and independent review.}
The same model audited and graded \DatasetITW{}, so shared errors
could inflate the 99.2\% support rate. Invalid rates (0.6\% here vs. 9.7\% on
\DatasetKF{}) span different audit pools and references. Selected expert notes
directly support 1 of 98 Critical routes (\S\,\ref{sec:production-human-review}).
Corpus-wide precision, grader accuracy, and independence from these notes
remain unknown. We did not separately penalize false positives or time expert
review per confirmed finding.
Severity indicates priority. Exploitability, repair difficulty,
and novelty require evidence for each flaw.

\pp{Usage accounting.}
We estimated missing \DatasetKF{} token counts from request
and response lengths using median token ratios and cache fractions from
completed requests. The largest estimated addition was about 2.3\% of
one model's recorded cost for one condition and round across papers.
Missing usage may exceed the 2.5\% added to affected reports.

\pp{Harness and effort.}
Each \DatasetKF{} model used the same harness across conditions.
Provider adapters and effort scales differ (\S\,\ref{sec:study}).
Higher settings often exhausted GLM's and MiniMax's context windows,
so both used \texttt{high}.
In the GPT-6.1 audits we examined, only playbook audits
used subagents and received a follow-up coverage check. The reported
evaluation did not test the coverage gain at equal inference cost
(\S\,\ref{sec:rq2-repeat}).

We are conducting follow-up experiments in which the baseline auditor is
required to continue auditing with a token budget approximately matched to
the playbook condition. These experiments examine whether additional
search narrows the observed performance gap.

\pp{Reusing the audit playbook.}
We will release the audit instructions. The black-box models that
proposed revisions, audited papers, and helped grade reports can give
different answers to the same input. We set the evidence requirements,
retained revisions that improved development scores
(\S\,\ref{sec:playbook-development}), and tested repeat detection with
fixed instructions (\S\,\ref{sec:rq2-repeat}).

\section{Related Work}
\label{sec:related-work}

\pp{LLM review of scientific papers.}
LLM review studies include human-feedback
comparisons~\cite{liang2024llmfeedback}, a randomized
trial~\cite{thakkar2025reviewfeedback}, and conference
pilots~\cite{biswas2026aaaipeerreview,jayaram2026pat}. Recent work reports
errors in cryptographic~\cite{woodruff2026gemini}, AI~\cite{bianchi2025toerr},
and mathematical papers~\cite{rivin2025euler}; SPOT finds low recall and
poor inter-run agreement~\cite{son2025spot}.
Agent scores vary across runs~\cite{bjarnason2026randomness};
pass\textasciicircum{}k measures consistency~\cite{yao2024taubench}.
We compare proof auditing with and without a human-aligned playbook using
three-run variation, union coverage, and repeat detection.

\pp{Proof grading and cryptographic benchmarks.}
LLMs grade natural-language
proofs~\cite{ma2025proofgrader,gonzalez2026qedbench}, detect reasoning
errors~\cite{jiang2024pedcot}, and verify proofs~\cite{naik2026verifyproofs},
though generative verifiers can be lenient~\cite{mahdavi2025verifiers}.
AICrypto evaluates cryptographic knowledge and proof
writing~\cite{wang2025aicrypto}; CryptoFormalEval detects protocol flaws
with a symbolic verifier~\cite{curaba2024cryptoformaleval}.
Our LLM grader scores reports against Answerbooks
aligned using $+P$ and $-P$ reports and targeted human
adjudication.

\pp{Playbooks, skills, and context engineering.}
EvoHunt evolves source-code auditing playbooks and reports
transfer across models~\cite{wang2026evohunt}; ACE evolves context
playbooks~\cite{zhang2025ace}; GEPA evolves prompts through
reflection~\cite{agrawal2025gepa}; and SkillOpt optimizes skills using a
held-out gate across models and harnesses~\cite{yang2026skillopt}.
SkillsBench finds benefits from curated skills but limited gains from
self-generated ones~\cite{li2026skillsbench}. We study how human-defined
criteria and Answerbook alignment guide playbook development for proof
auditing.

\pp{Security agents and benchmarks.}
Security agents target smart
contracts~\cite{zhou2025a1}, Android vulnerabilities~\cite{wang2025a2},
web applications~\cite{david2025mapta}, blockchain
attacks~\cite{wang2025txray}, and coordinated
exploits~\cite{fang2024hptsa}. Cybench~\cite{zhang2024cybench},
CVE-Bench~\cite{zhu2025cvebench},
CyberGym~\cite{wang2025cybergym,shi2026cybergym},
ExploitBench~\cite{lee2026exploitbench}, and
ExploitGym~\cite{wang2026exploitgym} use runtime oracles. Our audits need neither
oracle nor implementation.

\pp{Formal cryptographic assurance.}
CryptoVerif~\cite{blanchet2008cryptoverif},
EasyCrypt~\cite{barthe2011easycrypt}, and FCF~\cite{petcher2015fcf} check
formalized game-based proofs; theorem-proving agents also require
verifiers~\cite{yang2024leandojo,jiang2023dsp,alphaproof2024}.
Historical proof flaws motivate such
scrutiny~\cite{shoup2001oaep,bellare2001oaep,coron2002sflash,inoue2018ocb2}.
\toolname audits natural-language papers without prior formalization; its findings
require scientific review and provide no formal certificate.

\section{Conclusion}
\label{sec:conclusion}
\toolname uses expert feedback to guide audits and grade reports.
Its playbook improves coverage, rubric credit, and repeat detection for
five models on known-flaw papers and GPT-6.1 Sol in the wild. The planned
release of the playbook and known-flaw benchmark will support independent
evaluation.

\bibliographystyle{IEEEtran}
\bibliography{references}

\input{appendices}

\end{document}

%% file: figures/workflows/playbook-development-figure.tex
\begin{figure}[tb]
  \centering
  \includegraphics[width=\columnwidth]{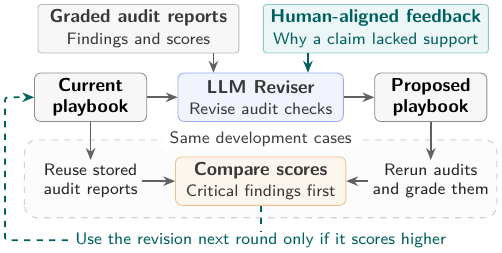}
  \caption{One revision round compares total credit on matched cases.
  Critical credit takes priority, then Medium and Low. A tie at all three levels
  retains the current playbook (Appendix~\ref{app:playbook-comparison}).}
  \label{fig:playbook-development}
\end{figure}

%% file: figures/workflows/audit-workflow-figure.tex
\begin{figure}[tb]
  \centering
  \includegraphics[width=\columnwidth]{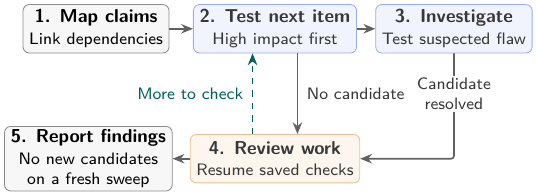}
  \caption{After investigating a candidate, the auditor
  resumes unfinished checks. It reports supported findings (step 5) once
  all candidates are resolved, material dependencies have been checked,
  and a fresh sweep after the last discovery finds no new independent
  candidate.}
  \label{fig:playbook-audit-workflow}
\end{figure}

%% file: playbook-checks-list.tex
\begin{enumerate}[label=\textbf{Example~\arabic*:},wide=0pt]
  \item \textbf{Distinguishing a false alarm from a proof flaw.}
    In a batch including NTS-KEM~\cite{albrecht2019ntskem}, an audit rejected
    a simulation without showing an adversary-observable difference before
    the exceptional (``bad'') event, though views need only match until then.
    Feedback also found a conditional-probability error. The Reviser required
    checking pre-event observations, consistency with earlier oracle answers,
    and probabilities separately: a valid simulation does not validate
    subsequent probability calculations.
  \item \textbf{Checking every security-relevant field.}
    Audits sometimes missed fields in composite objects. In the targeted
    UOV aggregate-signature case~\cite{chen2020mqaggregate}, the hash leaves
    the encoding tail unprotected, allowing altered signing histories to
    pass verification. The Reviser required tracing how each field is
    produced and protected, then testing changes to apparently unbound
    fields using adversary-computable values and checking verification,
    freshness (a new forgery under the game), and the winning condition.
  \Needspace{6\baselineskip}
  \item \textbf{Completing a probability argument.}
    In a batch including Trusted-HB~\cite{bringer2008trustedhb}, a report
    suggested repetition could violate a bound without a legal attempt or
    numerical comparison. The Reviser required an attempt with observable success,
    probability $p$, and permitted count $q$. For independent attempts with equal $p$,
    agents evaluate $1-(1-p)^q$ at the paper's parameters to test its bound.
  \Needspace{3\baselineskip}
  \item \textbf{Rechecking component guarantees.}
    Audits missed composition failures despite finding local defects. In
    the targeted AES masking case of Rivain and Prouff~\cite{rivain2010masking},
    intermediate values that appear safe individually can jointly reveal
    information about a secret.
    The Reviser required checking theorem premises on actual caller-supplied
    inputs, including their joint distribution. Missing premises support a
    proof gap; claiming a break additionally requires a concrete execution
    that violates the security property.
\end{enumerate}

%% file: figures/workflows/evaluation-workflow-figure.tex
\begin{figure}[tb]
  \centering
  \includegraphics[width=\columnwidth]{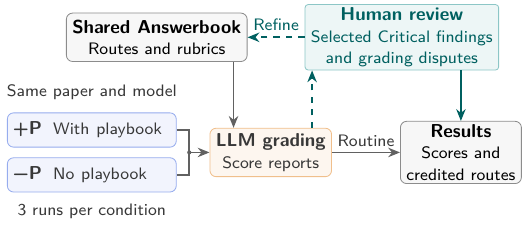}
  \caption{Evaluation on \DatasetKF{} and \DatasetITW{}.
  Both conditions share each paper's Answerbook. Humans review selected Critical
  findings and disputed credit. Changes to grading criteria trigger regrading of
  affected reports in both conditions.}
  \label{fig:pipeline}
\end{figure}

%% file: tables/benchmark-critical-results.tex
\begin{table}[tb]
  \centering\small
  \caption{Critical results on \DatasetKF{} (81 routes).}
  \label{tab:critical-overview}
  \setlength{\tabcolsep}{1pt}
  \begin{tabular*}{\columnwidth}{@{\extracolsep{\fill}}clrrr@{}}
    \toprule
    & Model & -P & +P & Gain [min, max] \\
    \midrule
    \multirow{5}{*}{\rotatebox[origin=c]{90}{\shortstack{\emph{Audit score}\\\emph{(0--1)}}}} & 6.1 Sol & 0.384$\pm$0.017 & \textbf{0.506}$\pm$0.036 & +0.121 [0.097, 0.145] \\
     & 6 Sol & 0.212$\pm$0.029 & 0.400$\pm$0.019 & +0.187 [0.160, 0.242] \\
     & 5.6 Sol & 0.196$\pm$0.018 & 0.469$\pm$0.034 & +0.273 [0.221, 0.312] \\
     & MiniMax & 0.146$\pm$0.045 & 0.320$\pm$0.022 & +0.173 [0.155, 0.206] \\
     & GLM & 0.104$\pm$0.013 & 0.255$\pm$0.025 & +0.151 [0.143, 0.166] \\
    \midrule
    \multirow{5}{*}{\rotatebox[origin=c]{90}{\shortstack{\emph{Coverage}\\\emph{(\%)}}}} & 6.1 Sol & 39.9$\pm$3.1 & 53.9$\pm$2.6 & +14.0 [12.3, 16.0] \\
     & 6 Sol & 23.5$\pm$3.3 & 44.4$\pm$2.1 & +21.0 [17.3, 27.2] \\
     & 5.6 Sol & 23.9$\pm$2.9 & \textbf{57.6}$\pm$3.6 & +33.7 [28.4, 39.5] \\
     & MiniMax & 18.1$\pm$6.3 & 44.0$\pm$1.9 & +25.9 [22.2, 30.9] \\
     & GLM & 14.4$\pm$0.7 & 35.8$\pm$4.3 & +21.4 [18.5, 25.9] \\
    \bottomrule
  \end{tabular*}
\end{table}

%% file: tables/baseline-repeatability.tex
\begin{table}[tb]
\centering
\caption{Of the Critical routes found in one run without the playbook,
what percentage did the model find in both other rounds?
p0 counts any credit. p50 also requires the route's median score
(\S\,\ref{sec:study}).}
\label{tab:baseline-repeatability}
\small
\setlength{\tabcolsep}{1pt}
\begin{tabular*}{\columnwidth}{@{\extracolsep{\fill}}llrrrr@{}}
\toprule
& & \multicolumn{2}{c}{p0 (\%)} & \multicolumn{2}{c}{p50 (\%)} \\
& Model & $-P\to+P$ & $\Delta$ (pp) & $-P\to+P$ & $\Delta$ (pp) \\
\midrule
\multirow{5}{*}{\DatasetKF{}} & 6.1 Sol & $80.7 \to 87.8$ & \textbf{+7.1} & $73.4 \to 87.8$ & \textbf{+14.4} \\
 & 6 Sol & $53.4 \to 86.6$ & \textbf{+33.2} & $28.3 \to 81.2$ & \textbf{+53.0} \\
 & 5.6 Sol & $47.0 \to 88.0$ & \textbf{+41.1} & $24.2 \to 70.5$ & \textbf{+46.3} \\
 & MiniMax & $52.8 \to 82.7$ & \textbf{+29.9} & $41.5 \to 69.6$ & \textbf{+28.0} \\
 & GLM & $34.3 \to 80.1$ & \textbf{+45.7} & $23.0 \to 37.4$ & \textbf{+14.4} \\
\midrule
\DatasetITW{} & 6.1 Sol & $68.1 \to 78.2$ & \textbf{+10.1} & $68.1 \to 78.2$ & \textbf{+10.1} \\
\bottomrule
\end{tabular*}
\end{table}

%% file: tables/three-run-cost-score.tex
\begin{table}[t]
\centering
\caption{One 6.1 Sol run plus each listed pair (\DatasetKF{}, $+P$).
Percent changes are relative to three 6.1 Sol runs.}
\label{tab:three-run-cost-score}
\small
\setlength{\tabcolsep}{1.5pt}
\begin{tabularx}{\columnwidth}{@{}Yr@{\,}r@{\hspace{6pt}}r@{\,}r@{}}
\toprule
Two additional runs & \multicolumn{2}{c}{Est. USD/paper} & \multicolumn{2}{c@{}}{Combined score} \\
\midrule
2$\times$ 6.1 Sol & 13.25 &  & 0.639 &  \\
2$\times$ GLM & 6.53 & $(-50.7\%)$ & 0.608 & $(-4.8\%)$ \\
GLM + MiniMax & 7.23 & $(-45.4\%)$ & 0.618 & $(-3.3\%)$ \\
2$\times$ MiniMax & 7.94 & $(-40.1\%)$ & 0.625 & $(-2.3\%)$ \\
6.1 Sol + GLM & 9.89 & $(-25.4\%)$ & 0.652 & $(+2.1\%)$ \\
6 Sol + GLM & 10.12 & $(-23.6\%)$ & 0.665 & $(+4.0\%)$ \\
6.1 Sol + MiniMax & 10.60 & $(-20.1\%)$ & 0.663 & $(+3.8\%)$ \\
6 Sol + MiniMax & 10.82 & $(-18.3\%)$ & 0.671 & $(+5.0\%)$ \\
\bottomrule
\end{tabularx}
\end{table}

%% file: tables/component-ablation.tex
\begin{table}[t]
\centering
\footnotesize
\caption{Component ablation on 12 equally weighted papers with 28 Critical routes. Combined results use each route's highest score across three runs.}
\label{tab:component-ablation}
\setlength{\tabcolsep}{1.5pt}
\begin{tabular*}{\columnwidth}{@{\extracolsep{\fill}}llrrr@{}}
\toprule
& Condition & Coverage (\%) & Score & Full credit (\%) \\
\midrule
\multirow{3}{*}{\shortstack[l]{\textit{Mean}\\\textit{per run}}} & No playbook & 34.5$\pm$6.9 & 0.322$\pm$0.061 & 21.5$\pm$5.2 \\
 & Protocol only & 51.6$\pm$1.6 & 0.489$\pm$0.009 & 37.3$\pm$4.7 \\
 & Full playbook & 53.9$\pm$12.1 & 0.485$\pm$0.069 & 35.6$\pm$3.5 \\
\midrule
\multirow{3}{*}{\shortstack[l]{\textit{Three runs}\\\textit{combined}}} & No playbook & 46.5 & 0.444 & 35.4 \\
 & Protocol only & 63.2 & 0.616 & 54.2 \\
 & Full playbook & 74.3 & 0.665 & 54.2 \\
\bottomrule
\end{tabular*}
\end{table}

%% file: tables/production-performance.tex
\begin{table}[tb]
  \centering\small
  \caption{\DatasetITW{} (GPT-6.1 Sol/xhigh). Three-run means; $\Delta$: paired gains [observed min, max] (coverage: pp).}
  \label{tab:production-performance}
  \setlength{\tabcolsep}{2pt}
  \begin{tabularx}{\columnwidth}{@{}lYrrr@{}}
    \toprule
    & Severity (routes) & $-P$ & $+P$ & $\Delta$ [min, max] \\
    \midrule
    \multirow{3}{*}{\shortstack[l]{\emph{Audit score}\\\emph{(0--1)}}} & Critical (98) & 0.433 & 0.652 & +0.220 [0.202, 0.230] \\
     & Medium (133) & 0.552 & 0.724 & +0.173 [0.138, 0.215] \\
     & Low (64) & 0.603 & 0.820 & +0.217 [0.170, 0.251] \\
    \midrule
    \multirow{3}{*}{\shortstack[l]{\emph{Coverage}\\\emph{(\%)}}} & Critical (98) & 46.6 & 71.1 & +24.5 [21.4, 26.5] \\
     & Medium (133) & 55.9 & 73.2 & +17.3 [13.5, 21.8] \\
     & Low (64) & 61.5 & 83.9 & +22.4 [18.8, 26.6] \\
    \bottomrule
  \end{tabularx}
\end{table}

%% file: tables/production-claims.tex
\begin{table}[tb]
  \centering\footnotesize
  \caption{LLM grading and delegated-review outcomes. Percentages use claim occurrences, including repeats: 1,609 with the playbook ($+P$), 1,049 without ($-P$).}
  \label{tab:production-claims}
  \setlength{\tabcolsep}{3pt}
  \begin{tabular}{@{}lrr@{}}
    \toprule
    Outcome & +P & -P \\
    \midrule
    Supported, credit-bearing claims & 668 (41.52\%) & 479 (45.66\%) \\
    Supported; no independent credit & 932 (57.92\%) & 558 (53.19\%) \\
    Judged invalid & 6 (0.37\%) & 9 (0.86\%) \\
    Assessed; unverified & 3 (0.19\%) & 3 (0.29\%) \\
    \bottomrule
  \end{tabular}
\end{table}

%% file: appendices.tex
\appendices

\section{Worked Scoring Example}
\label{app:rubric}

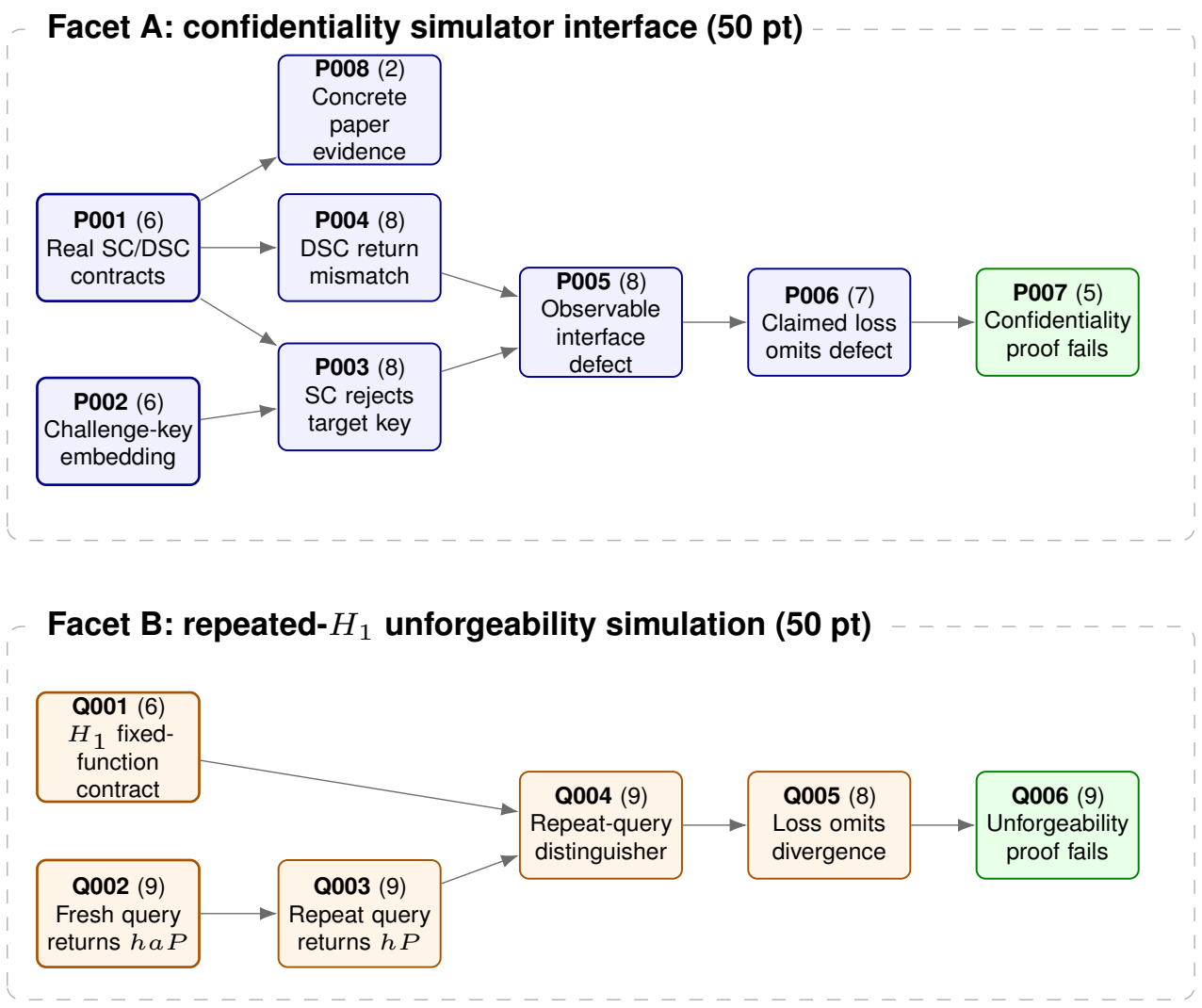
\begin{figure}[tb]
  \centering
  \resizebox{\columnwidth}{!}{%
    \input{figures/rubric-scoring-path.tex}
  }
  \caption{Historical scoring path with two cumulative 50-point facets.
  Nodes show positive weights, arrows indicate prerequisites (including the
  conjunctive gate), and Q labels abbreviate route-qualified identifiers.}
  \label{fig:rubric-case-study}
\end{figure}

Figure~\ref{fig:rubric-case-study} shows one declared path of a
historical training rubric outside \DatasetKF{}. Its two cumulative facets
each contribute 50 points: a confidentiality-interface mismatch and an
\(H_1\) repeat-query inconsistency in unforgeability simulation.
SC and DSC denote signcryption and designcryption
oracles; random oracles \(H_1,H_2\) must return consistent answers to
repeated inputs. A sibling path substitutes an \(H_2\) repeat-query witness
for the interface facet.

Suppose a report earns all 50 \(H_1\)-facet points and nodes
P001, P002, P004, and P008, but omits P003. The stored graph requires both
P003 and P004 for P005, so P005 and its descendants P006 and P007 earn zero.
This path scores \((50+6+6+8+2)/100=0.72\), with no extra deduction.
Declared alternative paths are scored independently; the route takes the
maximum, never combining incomplete paths.

Either adequately demonstrated observable interface mismatch
can invalidate the simulator, so the historical conjunctive gate is stricter
than logical sufficiency. The unchanged 0.72 measures completion of that
stored path, not minimum sufficient evidence for refutation.

\section{Supplementary Evaluation Results}
\label{app:extended}

\subsection{Sensitivity Checks}

\pp{Full rubric credit.}
Table~\ref{tab:reviewer-full-credit-summary} counts routes
receiving full credit for a declared path within one report. Gains remain
positive in every recorded round and model--dataset pair, but still depend
on rubric completeness and grading judgments.
\input{tables/full-credit.tex}

\pp{Attack-publication cohorts.}
Table~\ref{tab:appendix-sensitivity-summary} separates 20 cases
whose attacks predate all evaluated models' cutoffs from 22 whose attacks
were published after 30 April 2026. GPT-6.1's score gain is smaller in the
later group (0.037 versus 0.182); its p0 coverage gains are 5.9 versus 19.9
percentage points. Later publication does not establish absence of training
exposure, nor must shared exposure benefit audits equally with and without the playbook.

\pp{Reference snapshots.}
The same table compares the original reports under the initial
58-route and reviewed 81-route Critical references. Gains remain positive,
but their magnitude changes. The initial reference was already aligned:
these snapshots do not separate original published attacks from newly
admitted findings. The increase of 23 Critical routes is not a count of new
defects; scope, severity, and rubrics can change under retained route IDs.
\input{tables/sensitivity.tex}

\input{supplementary-results}

The artifact retains full tables, per-round results, the general
rubric specification, and supplemental analyses. The public release will
include the rubric JSON for the known-flaw Answerbook.

\section{Two Corrected Construction Flaws}
\label{app:corrected-constructions}
We audited Afshar and Goyal's MKFHE proceedings
paper~\cite{DBLP:conf/crypto/AfsharG26} in \DatasetITW{}. Two of the 98 Critical
routes concern its missing computations. The first full ePrint dates to
13 August 2026 (revised 8 October~\cite{afshar2026succinct}).

\pp{Mixed trees in distributed decryption.}
Independent partial-decryption executions sample their own
smudging values and hardcode them into evaluation circuits. Deterministic
evaluation does not synchronize these different circuits. For a shared
two-user ancestor, let \(A_1,A_2\) and \(B_1,B_2\) denote decryption
contributions from two separately generated parent ciphertexts. Base
correctness constrains the matched sums \(A_1+A_2\) and \(B_1+B_2\),
whereas the next split combines \(A_1\) with \(B_2\). The common-tree
argument therefore does not justify the mixed leaves used in final
decryption. After correcting the separate OTE argument-order typo,
conforming counterexamples preserve complete-ciphertext correctness while
mixed leaves fail: offsets that cancel within each parent need not cancel
across parents. This refutes the generic construction's correctness guarantee
without establishing failure of its particular external lattice
implementation. The October revision derives a common tree using deterministic FHE,
correlation-intractable hashing, and a trusted setup.

\pp{Colliding labels in threshold trapdoor hashing.}
Correctness requires exactly one decoded label to equal the
public hint \(e\). With honest identical predicates and perfect underlying
decryption, the decoder returns \((e,e')\) or \((e',e)\), depending on the
predicate value. Independent uniform one-bit labels collide with probability
\(1/2\); the decoded pair is then \((e,e)\), violating the unique-match
requirement for either predicate value. The first full ePrint already
requires \(e\ne e'\), which repairs the failure and preserves a one-bit hint.

\pp{Observed audit performance.}
Three GPT-6.1 Sol runs per condition received the following
normalized route scores, listed in recorded round order:
\begin{center}
\small
\begin{tabular}{@{}lcc@{}}
\toprule
Route & $+P$ & $-P$ \\
\midrule
Mixed trees & $(1,1,1)$ & $(1,0.6,0)$ \\
Colliding labels & $(1,1,1)$ & $(0,1,0)$ \\
\bottomrule
\end{tabular}
\end{center}
 Across three runs, GPT-6.1 Sol found both routes with and without the playbook.
The playbook reaches full rubric credit more consistently in these runs;
this does not establish exclusive discovery or independent correctness
confirmation.

These failures and corrections establish neither discovery priority nor
unknown defects in the current full version.
Excluding this paper leaves 120 papers and 96 Critical routes; mean
full-credit counts remain higher with the playbook, 54.00 versus 37.67 per
round. This does not re-audit the full version or validate other version
choices.

\section{Playbook Revision Comparisons}
\label{app:playbook-comparison}
We sum scores on matched cases, excluding pairs with missing scores,
grading failures, or unresolved human alignment. With no comparable cases, we keep
the playbook. Baseline reports are reused. For successful candidate runs, we grade
against the validation-start Answerbook and regrade baselines only if it changed.
Failed candidate runs score $(0,0,0)$ against stored baseline scores.

\section{Dataset and Reference Details}
\label{app:dataset-reference-details}

\pp{Historical selection and reference snapshots.}
From 26{,}387 ePrint records, we shortlist 2{,}501 candidate attack papers
using keywords and sentence patterns such as ``flaw,'' ``flawed,''
``refute,'' ``insecure,'' and ``break'' in abstracts, introductions, and
conclusions. An agent reads each candidate in full to identify what it
claims to break. A claimed break of a provable-security property forms a
candidate attack--victim pair. Source verification and human admission
retain 562 attack papers and 633 pairs. The historical split retains 619 pairs through
2025 and excludes 14 with 2026 attack papers.

\DatasetKF{} retains 20 of the original 36 test cases, with attacks
published in July--December 2025, and 22 of 29 newer candidates, whose
attack papers first appeared after 30 April 2026. Inadmissible pairs are
dropped without replacement. Retention requires complete three-round
reports from all five models with and without the playbook. In Critical/Medium/Low
order, the older group's reference changes from 34/49/5 (88 routes) to
47/41/10 (98); the newer group's changes from 24/67/16 (107) to
34/66/14 (114). Together they change from 58/116/21 (195) to
81/107/24 (212). Both snapshots grade the same reports, so these changes
reflect reference revision rather than changed auditor behavior.

\pp{Rubric generation and structure.}
An earlier OpenAI LLM (\texttt{gpt-5.6-sol}) receives both papers,
taxonomy, and output schemas in an isolated workspace. It checks scope,
re-derives the flaw in a sample answer, then constructs and self-checks a
rubric against that answer; the check may not weaken the rubric to make
the answer pass. Nodes are reasoning or evidence obligations with positive
weights \(w_v\), and edges are prerequisites. Each node earns
\(e_v\in[0,w_v]\); unmet prerequisites block credit without an extra edge
penalty. We score each sufficient path in \(\mathcal P_r\) separately:
\[
  s_p=\frac{\sum_{v\in p} e_v}{\sum_{v\in p} w_v},
  \qquad s_r=\max_{p\in\mathcal P_r}s_p .
\]
The 212 \DatasetKF{} route DAGs average 6.30 nodes and 5.92 edges
(ranges 3--24 and 2--32); the 295 \DatasetITW{} DAGs average 4.26 and
3.13 (3--6 and 2--5). Counts include every node and prerequisite once per
route, not selected-path sizes. Per-case sums average 31.81/29.86
nodes/edges over 42 \DatasetKF{} papers and 10.40/7.64 over all 121
\DatasetITW{} papers, including 23 with no reference route.
Appendix~\ref{app:rubric} gives a historical example.

\pp{Judgment states and revised credit.}
A finding may contain separately adjudicated claims. A supported claim
may receive no independent route credit when merged with another finding,
judged non-material, or excluded from the scoring scope. An investigated claim
that remains unverifiable is \emph{unverified} (serialized as
\texttt{uncertain}) and receives no credit; this differs from an unfinished
\texttt{pending} judgment. Confirmed correctness does not resolve a pending
independent-credit decision.
In P008 (\DatasetKF{}), delegated LLM review raised two unchanged baseline reports'
credit on one route from zero to 0.225 for partial dependency reasoning. A separate
claim in each remained invalid.

\pp{Estimated human effort.}
Table~\ref{tab:human-effort} estimates 220 person-hours for reference
preparation and technical review. We count five minutes per screened entry and one
hour per detailed review in the first three stages, including training and screening
beyond the evaluation set. Two reviewers spent about 1.5 hours each per
\DatasetITW{} paper across 13 papers (\S\,\ref{sec:production-human-review}). The
estimate excludes untimed discussions, playbook development, and earlier reviews.

\begin{table}[t]
  \centering
  \caption{Estimated human effort.}
  \label{tab:human-effort}
  \small
  \setlength{\tabcolsep}{2pt}
  \begin{tabular*}{\columnwidth}{@{\extracolsep{\fill}}lrrr@{}}
    \toprule
    Stage & Reviewed & Person-h/unit & Total person-h \\
    \midrule
    Training & 54~cases & 1 & 99 \\
    Validation & 42~cases & 1 & 45.6 \\
    Evaluation preparation & 31~cases & 1 & 34.6 \\
    \DatasetITW{} & 13~papers & 3 & 39 \\
    \midrule
    \multicolumn{3}{@{}l}{Total} & $\approx 220$ \\
    \bottomrule
  \end{tabular*}
\end{table}

\pp{In-the-wild collection and version checks.}
Of 662 publications, 521 had proceedings PDFs and 141 used public
ASIACRYPT and TCC preprints. The earliest-public-version cutoff left 125 papers,
of which 121 had in-scope formal claims (4{,}456 pages).
This first audit covered 71 CRYPTO, 20 ASIACRYPT, 13 EUROCRYPT, nine PKC, and eight
TCC papers. Three benchmark-victim positive controls predated the cutoff.
We prefer designated full texts, falling back to the most OCR words. We match
preprints by proceedings text overlap or, for preprint-only cases, by acceptance
metadata and title. Selected texts may predate the latest full version
(Appendix~\ref{app:corrected-constructions}).

\pp{Technical review and disclosure.}
We have sent disclosure emails highlighting nine human-reviewed concerns to the authors of seven in-the-wild papers and are preparing drafts for more papers.

In response to our disclosure, an author of one of these papers, published at CRYPTO 2026, confirmed a reported proof flaw by email, proposed fixes to mitigate it, and plans to update the preprint. We are awaiting further responses from contacted authors.

%% file: figures/rubric-scoring-path.tex
\begin{tikzpicture}[
  y=.88cm,
  font=\sffamily\tiny,
  rubricnode/.style={draw=black!60, rounded corners=2pt, align=center,
    text width=1.20cm, minimum height=0.82cm, fill=black!3,
    inner xsep=0.6pt, inner ysep=1.0pt},
  facetA/.style={rubricnode, fill=blue!6, draw=blue!55!black},
  facetB/.style={rubricnode, fill=orange!9, draw=orange!65!black},
  originnode/.style={line width=0.55pt},
  finalnode/.style={fill=green!9, draw=green!50!black},
  facetbox/.style={draw=black!30, rounded corners=3pt, dashed,
    line width=0.3pt},
  facetlabel/.style={font=\sffamily\scriptsize\bfseries, fill=white,
    inner xsep=2pt, inner ysep=0.5pt},
  rubedge/.style={-{Latex[length=1.35mm]}, line width=0.28pt,
    draw=black!58}
]
  \draw[facetbox] (-0.85,2.20) rectangle (8.25,6.65);
  \node[facetlabel, anchor=west] at (-0.62,6.65)
    {Facet A: confidentiality simulator interface (50 pt)};

  \node[facetA,originnode] (P001) at (0.0,4.75)
    {\textbf{P001} (6)\\ Real SC/DSC\\contracts};
  \node[facetA,originnode] (P002) at (0.0,3.15)
    {\textbf{P002} (6)\\ Challenge-key\\embedding};
  \node[facetA] (P003) at (1.85,3.45)
    {\textbf{P003} (8)\\ SC rejects\\target key};
  \node[facetA] (P004) at (1.85,4.75)
    {\textbf{P004} (8)\\ DSC return\\mismatch};
  \node[facetA] (P005) at (3.70,4.10)
    {\textbf{P005} (8)\\ Observable\\interface defect};
  \node[facetA] (P006) at (5.45,4.10)
    {\textbf{P006} (7)\\ Claimed loss\\omits defect};
  \node[facetA,finalnode] (P007) at (7.20,4.10)
    {\textbf{P007} (5)\\ Confidentiality\\proof fails};
  \node[facetA] (P008) at (1.85,5.95)
    {\textbf{P008} (2)\\ Concrete paper\\evidence};

  \draw[rubedge] (P001) -- (P003);
  \draw[rubedge] (P002) -- (P003);
  \draw[rubedge] (P001) -- (P004);
  \draw[rubedge] (P003) -- (P005);
  \draw[rubedge] (P004) -- (P005);
  \draw[rubedge] (P005) -- (P006);
  \draw[rubedge] (P006) -- (P007);
  \draw[rubedge] (P001) -- (P008);

  \draw[facetbox] (-0.85,-1.80) rectangle (8.25,1.45);
  \node[facetlabel, anchor=west] at (-0.62,1.45)
    {Facet B: repeated-$H_1$ unforgeability simulation (50 pt)};

  \node[facetB,originnode] (Q001) at (0.0,0.40)
    {\textbf{Q001} (6)\\ $H_1$ fixed-function\\contract};
  \node[facetB,originnode] (Q002) at (0.0,-1.05)
    {\textbf{Q002} (9)\\ Fresh query\\returns $haP$};
  \node[facetB] (Q003) at (1.85,-1.05)
    {\textbf{Q003} (9)\\ Repeat query\\returns $hP$};
  \node[facetB] (Q004) at (3.70,-0.28)
    {\textbf{Q004} (9)\\ Repeat-query\\distinguisher};
  \node[facetB] (Q005) at (5.45,-0.28)
    {\textbf{Q005} (8)\\ Loss omits\\divergence};
  \node[facetB,finalnode] (Q006) at (7.20,-0.28)
    {\textbf{Q006} (9)\\ Unforgeability\\proof fails};

  \draw[rubedge] (Q001) -- (Q004);
  \draw[rubedge] (Q002) -- (Q003);
  \draw[rubedge] (Q003) -- (Q004);
  \draw[rubedge] (Q004) -- (Q005);
  \draw[rubedge] (Q005) -- (Q006);
\end{tikzpicture}

%% file: tables/full-credit.tex
\begin{table}[tb]
  \centering
  \caption{Mean full-credit Critical routes per round (three rounds). One report must complete a declared scoring path. Human confirmation requires separate review.}
  \label{tab:reviewer-full-credit-summary}
  \footnotesize
  \setlength{\tabcolsep}{3pt}
  \begin{tabular*}{\columnwidth}{@{\extracolsep{\fill}}lrr@{}}
    \toprule
    Model & $-P$ & $+P$ \\
    \midrule
    \multicolumn{3}{@{}l}{\DatasetKF{} (81 Critical routes)} \\
    GPT-6.1 Sol & 26.67 & 38.00 \\
    GPT-5.6 Sol & 7.33 & 24.33 \\
    GPT-6 Sol & 9.67 & 23.67 \\
    GLM-5.3-Flash & 3.33 & 10.33 \\
    MiniMax-M3.1-Flash & 6.67 & 13.33 \\
    \midrule
    \multicolumn{3}{@{}l}{\DatasetITW{} (98 Critical routes)} \\
    GPT-6.1 Sol & 38.33 & 56.00 \\
    \bottomrule
  \end{tabular*}
\end{table}

%% file: tables/sensitivity.tex
\begin{table}[tb]
  \centering
  \caption{Critical overall-score gains ($+P$ minus $-P$), averaged over three rounds. Attack-date cohorts use reviewed grades. Snapshot comparisons use all 42 papers.}
  \label{tab:appendix-sensitivity-summary}
  \footnotesize
  \setlength{\tabcolsep}{2pt}
  \begin{tabular*}{\columnwidth}{@{\extracolsep{\fill}}lrrrr@{}}
    \toprule
    & \multicolumn{2}{c}{Attack date} & \multicolumn{2}{c}{Reference snapshot} \\
    \cmidrule(lr){2-3}\cmidrule(l){4-5}
    Model & Earlier & Later & Initial & Reviewed \\
    Critical routes & 47 & 34 & 58 & 81 \\
    \midrule
    GPT-5.6 Sol & +0.263 & +0.287 & +0.211 & +0.273 \\
    GPT-6 Sol & +0.232 & +0.126 & +0.211 & +0.187 \\
    GPT-6.1 Sol & +0.182 & +0.037 & +0.038 & +0.121 \\
    GLM-5.3-Flash & +0.170 & +0.124 & +0.127 & +0.151 \\
    MiniMax-M3.1-Flash & +0.176 & +0.169 & +0.247 & +0.173 \\
    \bottomrule
  \end{tabular*}
\end{table}

%% file: supplementary-results.tex
\subsection{Supplementary Result Details}
\label{app:rq-results-details}

\pp{Shared detections and changing composition.}
On \DatasetKF{}, shared-positive mean credit rises from
0.821 to 0.908 for GPT-5.6 Sol, 0.897 to 0.945 for GPT-6 Sol,
0.742 to 0.812 for GLM-5.3-Flash, and 0.814 to 0.881 for
MiniMax-M3.1-Flash. Of GPT-5.6's 53 paired route--round observations,
25 improve, four decline, and 24 tie. Only routes credited in both arms enter this
comparison.

MiniMax illustrates how expanding coverage changes the average.
Its 65 detections missed in the corresponding baseline run averaged
0.625 credit. With these included, mean credit across detections fell from 0.807 to
0.726 despite higher credit on matched detections.

GPT-6.1's detected sets contain 97 baseline and 131 playbook
route--round observations: 90 shared, seven baseline-only, and 41
playbook-only. The additional observations average 0.866 credit, below
the 0.970 mean on shared detections with the playbook. Adding these observations
lowers mean credit among detections from 0.963 to 0.938 but raises overall score on
the fixed reference set. The change reflects which flaws were detected.

GPT-6.1's mean \DatasetKF{} p50 coverage is 43.67 with
versus 31.00 without the playbook ($+12.7$ routes).

\pp{Cumulative coverage and whole-set variation.}
For GPT-5.6 Sol, one, two, and three selected playbook rounds
cover 46.67 [45, 50], 53.67 [52, 55], and 57 Critical routes on
\DatasetKF{}; baseline counts are 19.33 [18, 22], 27.33 [23, 30], and 33.
Brackets give subset minima and maxima, not confidence intervals.
Across models, 17--36 routes with the playbook and 4--26 without it
receive credit in all three rounds. These comparisons budget audit
counts, not inference cost.

For GPT-6.1 on \DatasetKF{}, the number of routes found in
every run rose from 26 to 33, while the number found at least once rose
from 39 to 55 (Figure~\ref{fig:metric-persistence}c). The share found
every time therefore fell from 66.7\% to 60.0\%, and mean pairwise
Jaccard similarity fell from 76.4\% to
71.4\%. Both overlap ratios improve for the other four models.
Coverage-count standard deviation decreases for three models but increases
for GPT-5.6 and GLM. On \DatasetITW{}, the all-three intersection grows
from 31 to 46 routes and the union from 63 to 95; intersection/union
changes from 49.2\% to 48.4\%, and mean pairwise Jaccard from 62.2\% to
62.1\%. Coverage is $46.6\%\pm3.1\%$ without and $71.1\%\pm4.8\%$ with
the playbook (mean $\pm$ sample standard deviation over three rounds).
Larger recurring sets need not imply greater relative overlap
or lower count variation.

On \DatasetITW{}, paired score gain averages 0.2196
[0.2020, 0.2296] and net coverage gain 24 [21, 26] routes; brackets
give round ranges. Mean within-round shared coverage is 39 routes.
At p50, 69.67 versus 44.67 routes qualify, a net gain of 25.0;
this filters the same reports rather than generating new discoveries.
Three-round unions contain 35 playbook-only, three baseline-only,
and 60 shared routes.

On \DatasetKF{}, replacing two of three GPT-6.1 playbook audits with
MiniMax raises union coverage from 55 to 59.67 routes but lowers combined score from
0.639 to 0.625.

\pp{Outcome labels and residual uncertainty.}
For \DatasetKF{}, the grading records contain 8,698 claim occurrences:
7,725 supported, 840 judged invalid, and 133 unverified, with no
pending grading decisions. The LLM grader applied the criteria and
decisions we recorded in each Answerbook, and we reviewed selected new
or disputed claims (\S\S\,\ref{sec:alignment}--\ref{sec:grading}). Counts include
repeats. The 5,221 supported claims without independent route credit illustrate why
zero credit cannot establish a false positive.

For \DatasetITW{}, credit-bearing supported claims combine
the historical \emph{accepted} and \emph{partial} labels. The table groups these
inconsistently used labels; the records retain both. The 1,490 supported
noncredit occurrences are not exhaustively classified as non-material,
duplicate, or out of scope by their free-text exclusions. Eight reports
with no issues remain in the 726-report denominator. Their summaries document checks
with no reportable finding; they cannot certify exhaustive auditing or the absence of
flaws.

All 15 claim occurrences judged invalid in \DatasetITW{} carry
the grading label \texttt{ocr\_transcription\_error} and span seven papers.
Recorded rationales cite original-PDF contradictions involving
lost overbars, inverse superscripts, or deletion marks. This is the stored
diagnosis, not an independent PDF readjudication or an OCR-intervention
result. Visual capability or an asserted check does not establish
successful PDF access or interpretation.
Of six unverified occurrences with final no-credit decisions,
five concern an unresolved distributional interpretation in one paper;
one depends on omitted eligibility and state-update rules in another.
Final dispositions do not resolve these scientific uncertainties.

The 16 findings span eight technical areas and came from first-round
playbook reports (6 Critical, 9 Medium, 1 Low routes). A separate LLM analyst per
finding read the report and comments blind to the grade. We matched the assertions
to graded claims and routes. Notes supported 3 conclusions (1 with the same
mechanism, 2 in context without deriving it), raised 2 broader concerns, qualified
3 findings, and gave no claim-specific judgment for 8. For a second Critical route,
a note questioned a theorem assumption without identifying the defining consistency
condition. Earlier review activity has not been reconciled with this count.

\pp{Ablation panel selection.}
The twelve-paper panel takes the first two papers in the
fixed hash order within each of six strata. Strata cross
historical or post-cutoff attack sources with Critical-route composition:
construction-only, mixed, or non-construction. In that order, the
historical pairs are P038/P007, P008/P016, and P022/P015; the post-cutoff
pairs are P021/P003, P042/P018, and P034/P027. The ranking uses paper
identifiers rather than scores. The panel contains 28
Critical reference routes. Equal representation of strata differs from
the benchmark's natural distribution.